\documentclass{aa}
\usepackage{graphicx}
\usepackage{txfonts}
\usepackage{natbib}
\bibpunct{(}{)}{;}{a}{}{,} 
\newcommand{\mincir}{\raise -2.truept\hbox{\rlap{\hbox{$\sim$}}\raise5.truept
\hbox{$<$}\ }}
\newcommand{\magcir}{\raise -2.truept\hbox{\rlap{\hbox{$\sim$}}\raise5.truept
\hbox{$>$}\ }}
\newcommand{\siml}{\raise -2.truept\hbox{\rlap{\hbox{$\sim$}}\raise5.truept
\hbox{$<$}\ }}
\newcommand{\simg}{\raise -2.truept\hbox{\rlap{\hbox{$\sim$}}\raise5.truept
\hbox{$>$}\ }}
\newcommand{\be}{\begin{equation}}
\newcommand{\ee}{\end{equation}}
\newcommand{\ba}{\begin{eqnarray}}
\newcommand{\ea}{\end{eqnarray}}

\newcommand {\ks} {km~s$^{-1} \;$}
\newcommand {\kss} {km~s$^{-1}$}

\newcommand {\mqua} {$\times 10^{14}\;M_{\odot} \;$}
\newcommand {\mquaa} {$\times 10^{14}\;M_{\odot}$}
\newcommand {\mtre} {$\times 10^{13}\;M_{\odot} \;$}

\newcommand {\mdod} {$\times 10^{12}\;M_{\odot} \;$}

\newcommand{\degree}{\ensuremath{\mathrm{^\circ}}}
\newcommand{\arcm}{\ensuremath{\mathrm{^\prime}\;}}
\newcommand{\arcs}{\ensuremath{\arcmm\hskip -0.1em\arcmm \;}}
\newcommand{\arcmm}{\ensuremath{\mathrm{^\prime}}}
\newcommand{\arcss}{\ensuremath{\arcmm\hskip -0.1em\arcmm}}

\newcommand{\dotsec}{\,\rlap{\hbox{$\mathrm{^s}$}}{\hbox{$.$}}\,}

\newcommand{\emi}{${\it GRS0917+75} \;$}
\newcommand{\emii}{{\it GRS0917+75}}
\newcommand{\ma}{\mathrm{mag}}
\newcommand{\maa}{\mathrm{mag\,arcsec^{-2}}}

\begin{document}
   \title{The giant radio source 0917+75: Origin and properties}  

   \author{G. Giovannini\inst{1,2}, 
   N. Biava\inst{1,3},
   M. Girardi\inst{4,5},
   W. Boschin\inst{6,7,8},
   R. Barrena\inst{7},
   A. Bonafede\inst{1,2},
   L. Feretti\inst{1},
   C. Ferrari\inst{9},
   F. Govoni\inst{10},
   M. Iacobelli\inst{11},
   M. Murgia\inst{10},
   E. Orru'\inst{11},
   R. Pizzo\inst{11},
   V. Vacca\inst{10}
}


\institute{INAF - Istituto di Radioastronomia, via Gobetti 101, 40129 Bologna, Italy \email{gabriele.giovannini@inaf.it}
\and Dipartimento di Fisica e Astronomia dell'Universit\'a degli studi di Bologna, via Gobetti 93/2, 40129 Bologna, Italy
\and Thuringer Landessternwarte, Sternwarte 5, 07778 Tautenburg, Germany
\and Dipartimento di Fisica dell'Universit\`a degli Studi di Trieste -
Sezione di Astronomia, via Tiepolo 11, I-34143 Trieste, Italy 
\and INAF - Osservatorio Astronomico di Trieste, via Tiepolo 11,
I-34143 Trieste, Italy
\and Fundaci\'on Galileo Galilei - INAF (Telescopio Nazionale
  Galileo), Rambla Jos\'e Ana Fern\'andez Perez 7, E-38712 Bre\~na
  Baja (La Palma), Canary Islands, Spain
\and Instituto de Astrof\'{\i}sica de Canarias, C/V\'{\i}a L\'actea
s/n, E-38205 La Laguna (Tenerife), Canary Islands, Spain
\and Departamento de Astrof\'{\i}sica, Univ. de La Laguna, Av. del
Astrof\'{\i}sico Francisco S\'anchez s/n, E-38205 La Laguna
(Tenerife), Spain
\and Universit\'e C\^ote d'Azur, Observatoire de la C\^ote d'Azur, CNRS, Laboratoire Lagrange, France 
\and INAF - Osservatorio Astronomico di Cagliari, Via della Scienza 5 - 09047 Selargius (CA), Italy
\and ASTRON, Postbus 2, 7990, AA, Dwingeloo, The Netherlands }

\date{Received  / Accepted }

\abstract
{Several author have studied the giant radio source \emi, but its origin remains unclear. }
{This source is unusual, because of its large size and its location outside a rich cluster of galaxies. We aim to understand and discuss the properties and nature of this source and its connection to the environment.}
{We conducted optical observations to obtain new spectroscopic data. We also acquired a LOFAR image at 144 MHz to derive information at low radio frequency. Moreover, we performed a new analysis of archival VLA data in the L and C bands for a multifrequency study of the source properties in the radio band.}
{From the observational data, we classify \emi as a giant radio galaxy with a size of 1.5 Mpc and an estimated age of about 100 Myr. The optical parent  galaxy is a bright low-excitation radio galaxy, the brightest member of a very poor group belonging to a large supercluster. \emi is a peculiar low-power Fanaroff-Riley Class I giant radio galaxy with a bright central emission but no jet-like features.} 
{The existence of giant radio galaxies such as \emi could explain the origin of relativistic particles and magnetic fields in low-density environments. }

\keywords{Galaxies: clusters: general -- Galaxies: kinematics and dynamics -- Galaxies: radio -- Giant radio galaxies}
\titlerunning{The giant radio source 0917+75} 
\authorrunning{Giovannini et al.}

\maketitle
\nolinenumbers

\section{Introduction}
\label{intro}

The giant radio source \object{0917+75} (hereafter \emii) is an elongated diffuse source discovered by \citet{dewdney1991} and studied by \citet{harris1993}. They found that this source appeared to be associated with two galaxies at redshift $z\sim 0.125$ 
\citep[specifically $z = 0.1241$ and $z = 0.1259$;][]{dewdney1991}
and therefore lies at the same redshift as the adjacent Rood 27 supercluster \citep[]{rood1976}: \object{Abell~762} with $z=0.1332$; \object{Abell~786} with $z=0.1241$; and \object{Abell~787} with $z=0.1352$ \citep{struble1987}, hereafter A762, A786, and A787, respectively. \citet{chen2008}
observed this source and a similar target (1401-33) with XMM-Newton. The absence of detected X-ray emission from Inverse Compton (IC) between synchrotron relativistic electrons and low-energy Cosmic Microwave Background (CMB) photons allowed the authors to estimate a lower limit on the magnetic field strength (0.81 {\textmu}G, 3$\sigma$ level).
\citet{harris1993} considered this source to be the remnant of  an old radio galaxy, while \citet{giovannini2015}, discussing the properties of this source from VLA observations \citep[]{giovannini2000}, suggested that it could be an inherent feature of the filamentary structure of the Rood 27 supercluster, or the remnant of a giant radio galaxy.

This source is peculiar because it is extended and diffuse with a steep radio spectrum similar to radio halos or relics in clusters of galaxies; however, it appears to be associated with a poor group without X-ray emission, while radio halos and relics occur in rich clusters with extended diffuse X-ray emission \citep[see e.g.,][]{feretti2012}.

Our current knowledge indicates that the optical field around \emi is very complex. The galaxy near the peak of radio emission is a bright elliptical galaxy ($r$-band
magnitude $r=16.58$) with a Sersic index $n=3.68\pm0.01$ \citep[Legacy Survey DR10 catalog;][]{dey2019}. To the southeast, at a
distance of less than $0.5$\arcmm, there is a disk galaxy that is one
mag fainter ($r=17.59$, $n=1.74\pm0.01$). 
These two galaxies are the brightest galaxies of a group that we identify from the available data, and we refer to them as BGG1 and BGG2, respectively.

The nearest Abell cluster, at $\sim~25$\arcm from the peak of the \emi radio emission,  is A786. It is a poor cluster \citep[richness $=0$;][]{abell1958} with low X-ray luminosity \citep[$L_{\rm X}=0.88\times 10^{44}$ erg s$^{-1}$ in the 0.1-2.4 keV band, from ROSAT HRI; ][]{boehringer2000}. According to the reanalysis of  \citet{chowmartinez2014}, the large-scale structure of the region is better described by two superclusters rather than the only Rood~27 supercluster. A786 is one of 13 galaxy clusters that form the rich supercluster SCL245 at $z_{\rm SCL}=0.123$. This supercluster extends over a very large region of the sky, approximately $10\times10$ degrees$^2$, but within a small redshift range, $\Delta z=0.012$.  This indicates that SCL245 lies mostly in the plane of the sky. The SCL203 supercluster lies in the background of the \emi radio emission and consists of nine clusters, including  A762 and A787. SCL203 lies in a region of approximately $2.5\times7.5$ degrees$^2$ and extends along the line of sight, from $z=0.1332$ to $z=0.1949$ ($z_{\rm SCL}=0.142$).

We used new and archival radio and optical data to investigate the nature and properties of this source and its connection to A786 and the supercluster environment. In Sect.~2, we analyze a redshift catalog of the \emi region based on new data from the Italian Telescopio Nazionale {\em Galileo} (TNG) combined with Dark Energy Spectroscopic Instrument (DESI) Data Release 1 (DR1).
In Sect.~3, we present new LOw Frequency ARray (LOFAR) and archival Very Large Array  (VLA) radio data. We discuss the results  in Sect.~4 and present our conclusions in Sect.~5. 

In this paper, we adopt $H_0=70$ km s$^{-1}$ Mpc$^{-1}$ in a flat
cosmology with $\Omega_0=0.3$ and $\Omega_{\Lambda}=0.7$. In this cosmology, 1\arcm corresponds to 133~kpc at the redshift of the group associated with  the two galaxies related to \emi ($z = 0.1240$, see Sect.~\ref{gru}); we use this convention throughout the paper to compute physical sizes and projected distances. The luminosity distance is 580 Mpc. Unless otherwise stated, errors are reported at the  68\% confidence level (c.l.). The velocities we derive for galaxies are
line-of-sight velocities determined from the redshift, $V=cz$.

\section{Optical data: Analysis and results}

\subsection{New redshift data and galaxy catalog}
\label{cat}

We conducted optical observations to obtain new spectroscopic data with the TNG in the \emi region, specifically within $25\times15$ arcmin$^2$. 
We performed multi-object spectroscopy (MOS) and long-slit spectroscopic
observations of galaxies in the \emi field at the TNG in April and May 2021, as well as during several observing periods from December 2022 to March 2024. For these observations, we used the Device Optimized for Low Resolution Spectroscopy (DOLoRes) with the low resolution blue grism.  This grism enables observation of all relevant spectral features at the group redshift and provides sufficient resolution for accurate galaxy redshift measurements. In total, we obtained spectra for 84 galaxies with exposure times ranging from 900~s to 1800~s.  We performed spectral reduction and radial velocity estimation using the standard Image Reduction and Analysis Facility
(IRAF)\footnote{IRAF is distributed by the National Optical Astronomy Observatories, which are operated by the Association of Universities  for Research in Astronomy, Inc., under a cooperative agreement with the National Science Foundation.} tasks and the cross-correlation technique \citep{tonry1979}. Based on our experience with DOLoRes in spectroscopic mode, the nominal velocity errors provided by the cross-correlation technique should be multiplied by a factor of 2.0 for long-slit spectra and 2.5 for MOS spectra \citep[e.g.,][]{girardi2022}. The median uncertainty in velocity measurements is 110~\kss.

The DESI-DR1 survey recently reported redshifts for many galaxies in the \emi region \citep{desi2024, desidr12025}. In total, 37 of these galaxies have redshifts in common with the TNG sample. These 37 galaxies allowed us to verify whether we could combine our TNG redshifts. We measured a systematic offset of approximately -200 km/s between the TNG MOS $cz$ and DESI $cz$ measurements. By analyzing the acquisition frames taken during the MOS observations, we found  that this offset is due to technical issues with the instrument and telescope setup. We corrected all MOS $cz$ measures for this offset to match those measured by DESI. In general, when both TNG and DESI
redshifts were available, we used DESI redshifts because of their smaller uncertainty.  For 46 galaxies in the \emi region, only TNG redshifts were available. 
Although the TNG data are important for completing the sampling of galaxies in the \emi region,  
the DESI data are necessary to map A786 and the surrounding regions. For A786, we obtained the long-slit spectrum of the Brightest Cluster Galaxy (BCG) from TNG observations, measuring  a redshift of $z=0.1253\pm0.0001$.  The  radial velocities $V=cz$ of these 47 new TNG redshifts are electronically
published in Table~\ref{tabcatalog}, available at the Strasbourg astronomical Data Center (CDS).

\begin{table}[ht]
\centering
    \caption[]{New TNG spectroscopic data for galaxies (extract).}
\label{tabcatalog}
           $
           \begin{array}{l c  c c }
            \hline
            \noalign{\smallskip}

\mathrm{ID} & \rm{R.A.,\ Dec.}& \rm{V} & r\\
            & {\rm h\ m\ s},\ \degree\ \arcmm\ \arcss& \rm{km\ s}^{-1} & \rm{mag}\\

         \hline
         \noalign{\smallskip}
1&09~28~01.69,+74~48~17.0&37561\pm42&16.77\\
2&09~21~20.89,+75~03~21.9&37181\pm65&18.47\\  
  \noalign{\smallskip}
  \hline
           \end{array}
$\tablefoot{The full table is available at the CDS. Only a portion is shown here to illustrate its form and
             content. Col.~1: Running ID of the galaxies in the
             sample. The BCG of Abell~786 is denoted ID~1. Col. 2: R.A. and Dec. (J2000).  Col. 3:
             Radial velocity, $V=cz$. Col.~4: $r$-band magnitude from Pan-STARRS.}
\end{table}

\subsection{Detection and properties of the \emi group}
\label{gru}

\begin{figure} 
\centering
\resizebox{\hsize}{!}{\includegraphics{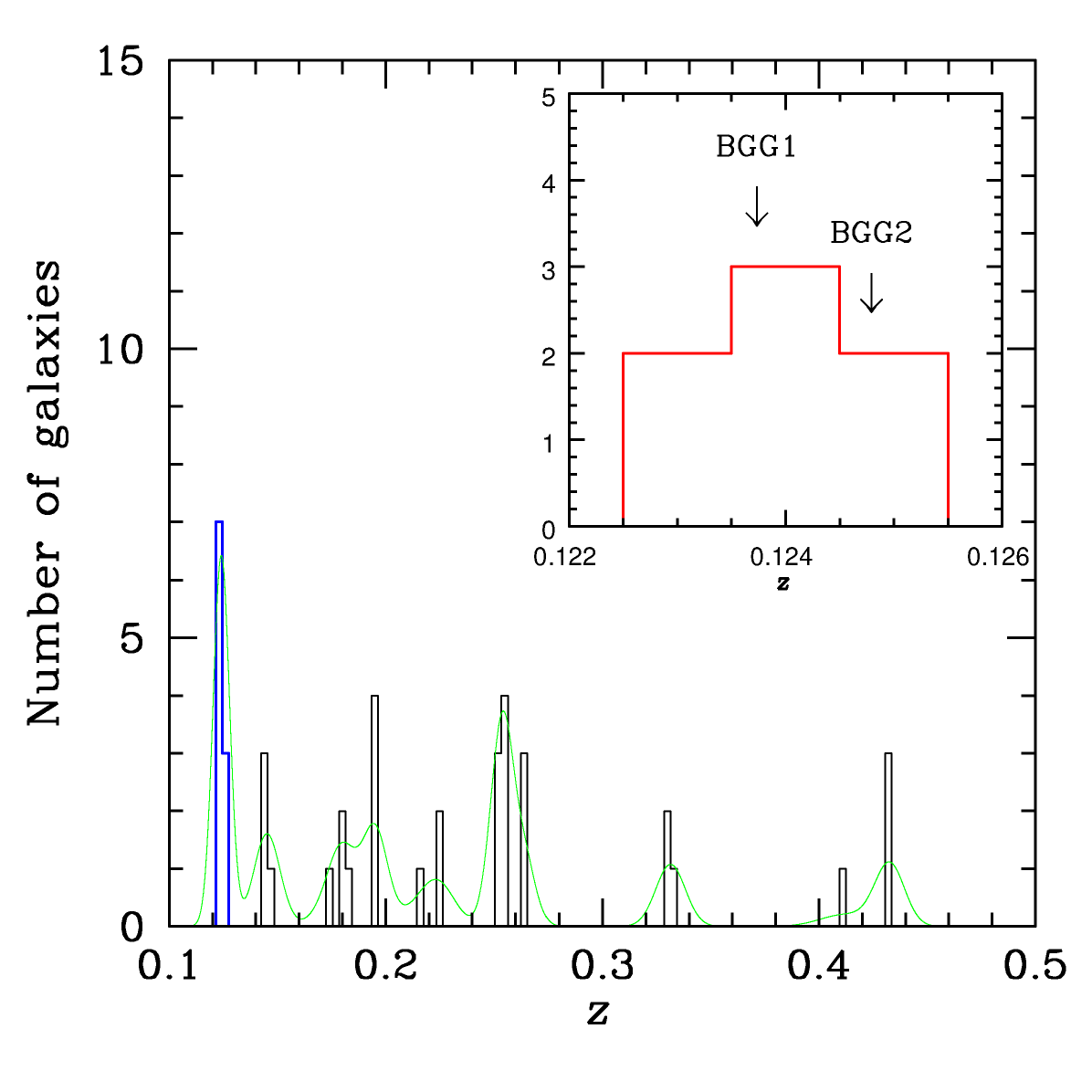}}
\caption
{Distribution of spectroscopic redshift of galaxies. The histogram shows the 47 galaxies with spectroscopic redshift within the 5\arcm radius around BGG1, while the green line shows the  density reconstruction from the 1D-DEDICA
  analysis in arbitrary y-units.  The blue histogram corresponds to the peak of ten galaxies surrounding BGG1; seven of these are clustered in the sky and selected as \emi group members (see text and Fig.~\ref{compact}). The inset shows their redshift distribution and also marks the redshifts of BGG1 and BGG2.}
\label{fighisto}
\end{figure}

\begin{figure*}
\centering
\resizebox{\hsize}{!}{\includegraphics{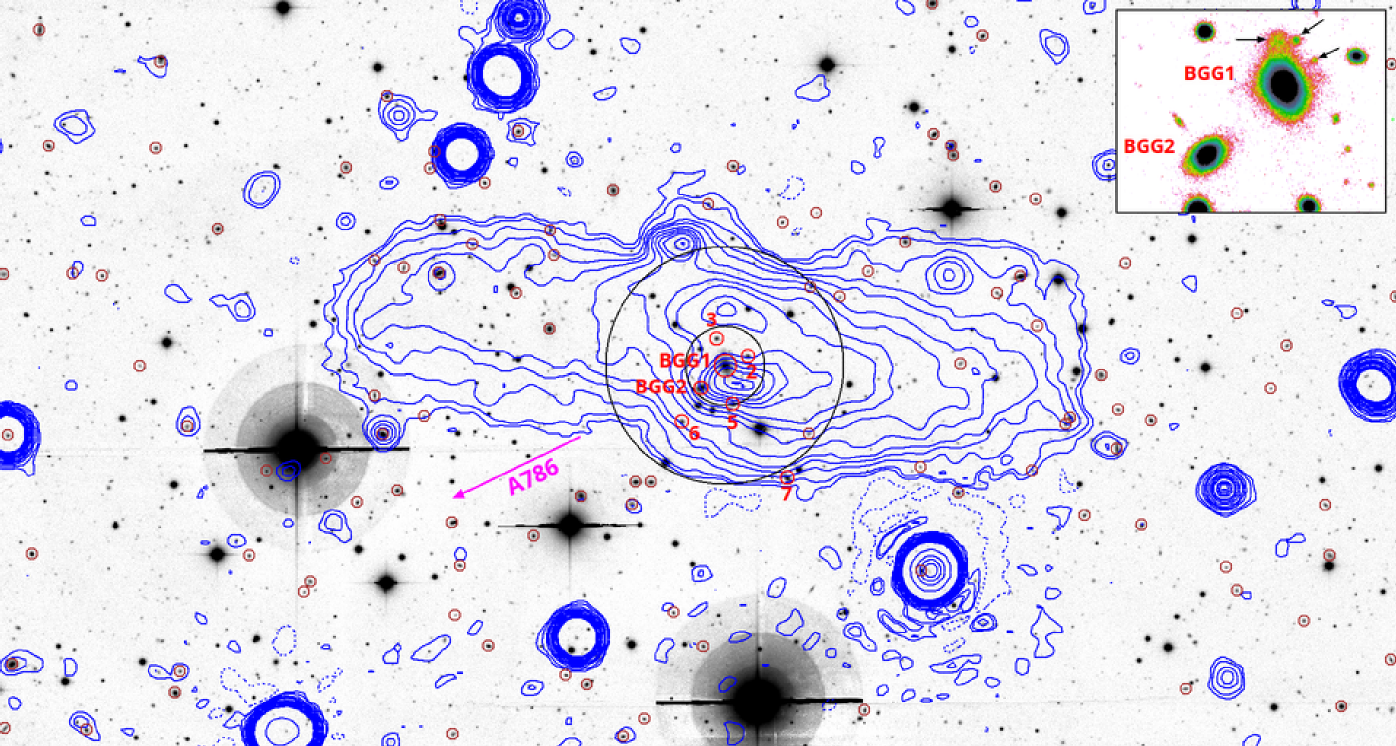}}
\caption
    {Zoomed region of a wide-field Isaac Newton Telescope (INT) $R$-band image with overlaid small circles marking galaxies with spectroscopic redshifts. Red circles and labels indicate members of the \emi group, while brown circles indicate non-members based on redshift and/or position (see text). Blue contours are from a LOFAR image with HPBW = 20 arcsec. Levels are -0.3, 0.3, 0.5, 1, 1.5, 2, 3, 4, 5, 6, 8, 9, 10, 11, 12, 100, 300, 500, and 700 mJy/beam (see Section 3.1). The inner large black circle has a radius of approximately 0.6\arcmm, and the outer circle is three times larger.  These circles are centered on the brightest galaxy group (BGG1) and include the isolation region of the fiducial compact group (see text). The diameter of the outer circle corresponds to 480 kpc at the group redshift ($z = 0.1240$). The inset (top right) shows the details of BGG1 and the surrounding region. BGG2 and other group galaxies are visible, as well as possible structures of BGG1 (highlighted by small black arrows).}
\label{compact}
\end{figure*}

We selected galaxies within a 5\arcm radius around BGG1 ($\sim 0.67$ Mpc) to investigate galaxies possibly associated with the \emi radio source and to study its surrounding medium. This sample consists of 47 galaxies: 22 measured only by TNG, eight only by DESI, and 17 by both. We analyzed the redshift distribution of this sample using the one-dimensional adaptive-kernel Density Estimation for the Identification of Clusters of galaxies \citet[][DEDICA;]{pisani1993}. We detect a peak of ten galaxies at $z\sim0.124$ (see Fig.~\ref{fighisto}).  These galaxies show a clear overdensity in their distribution on the
plane of the sky, as seen in Fig.~\ref{compact}, confirming the existence of a real group whose two brightest galaxies are BGG1 and BGG2. When 2D-DEDICA is applied to the galaxy positions in the
sky, we find a 2D  peak of galaxies with a significance level greater than 99\%. It contains the seven central galaxies, which are listed in Table~\ref{tabgroup}. We also applied 3D-DEDICA \citep{pisani1996} to the initial sample of 47 galaxies and find a peak containing the same seven galaxies, with a significance greater than 99.99\% c.l. In the following, we refer to these seven galaxies as the \emi group. Figure~\ref{compact} shows the \emi group (grayscale image) and the LOFAR radio image with HPBW = 20 arcsec (blue contours) presented in Sect. 3.1 (see the caption for more details).

\begin{table}[ht]
    \caption[]{Galaxies of the \emi group.}
\label{tabgroup}
           $
           \begin{array}{l c  c c }
            \hline
            \noalign{\smallskip}

\mathrm{ID} & \rm{R.A.,\ Dec.}& \rm{V} & r\\
            & {\rm h\ m\ s},\ \degree\ \arcmm\ \arcss& \rm{km\ s}^{-1} & \rm{mag}\\

         \hline
         \noalign{\smallskip}
1~{\rm (BGG1)}&09~22~09.57,+74~59~50.5&37095\pm 6\phantom{11}&16.58\\
2 &09~22~04.32,+74~59~59.2&37337\pm 7\phantom{11}&19.48\\
3 &09~22~11.70,+75~00~15.3&36989\pm  115&19.55\\
4~{\rm (BGG2)}&09~22~15.21,+74~59~30.8&37412\pm 85\phantom{1}&17.59\\ 
5 &09~22~07.72,+74~59~16.5&37246\pm 10\phantom{1}&18.78\\ 
6 &09~22~19.60,+74~59~00.0&37179\pm 3\phantom{11}&19.54\\
7 &09~21~54.98,+74~58~09.4&36971\pm 6\phantom{11}&18.76\\

  \noalign{\smallskip}
  \hline
           \end{array}
$\tablefoot{Galaxies are ordered by their distance from BGG1. Col.~1: Running ID for galaxies in the sample. Col. ~2: R.A. and Dec. (J2000). Col.~3: Radial velocity, $V=cz$; redshifts of IDs~3 and 4 are from TNG, those of the others are from DESI-DR1. Col.~4: $r$-band magnitude from Pan-STARRS.}

\end{table}

\begin{table}
\centering
        \caption{Global properties of the \emi group.}
         \label{tabv}
         $
         \begin{array}{c c c  c}
            \hline
            \noalign{\smallskip}

{\rm R.A., Dec.(J2000)}&z&\sigma_{V}&M_{200,L}\\
\mathrm{h\ m\ s, \degree\ \arcmm\ \arcs}& &\mathrm{km\ s^{-1}}&10^{13}M_{\odot}\\
         \hline
         \noalign{\smallskip}
09\,22\,09.57,+74\,59\,50.5&0.1240&156_{-32}^{+33}&2.1\pm 0.3\\
              \noalign{\smallskip}
            \hline
         \end{array}
         $\tablefoot{Col.~1: Group center, assumed to be
the position of BGG1. Col.~2: Mean redshift, $z$. Col.~3: 
Velocity dispersion, $\sigma_V$, and its uncertainty.
Cols.~4 and 5:  $M_{200}$ mass value derived from the optical luminosity,  $M_{200,L}$. Alternative, lower-mass estimates are discussed in the text.}
         \end{table}

We computed the mean line-of-sight velocity $\left<{V}\right>=37\,175\pm$67 \kss using the biweight estimator; the group redshift is $z=0.1240\pm0.0002$.  We estimated the velocity dispersion,
$\sigma_{V}$, by applying the cosmological correction and the standard correction for velocity errors \citep{danese1980}. We obtain $\sigma_{V}=156_{-32}^{+33}$ \kss, with errors estimated by the bootstrap method. 
Table~\ref{tabv} summarizes the main properties of the \emi group.

For the group mass, we provide two estimates based on different proxies: the group's optical luminosity 
and the velocity dispersion of member galaxies.
After correcting $r$ magnitudes of group galaxies for extinction using the values listed by  the NASA/IPAC Extragalactic database (NED)\footnote{Operated by the Jet Propulsion Laboratory, California Institute of Technology, under contract with NASA.} and applying the K correction\footnote{http://kcor.sai.msu.ru/} we obtain a group luminosity of $L_{\rm r}\sim 1.22x10^{11} L_{\sun}$. Using the  $M_{200}$-$L_{\rm opt}$ relation from Eq.~6 of \citet{popesso2007}, also estimated in the $r$ band, we derive $M_{200,L}\sim 2.1\pm0.3$ \mtre ($R_{200,L}=0.55$ Mpc). The reported uncertainty reflects the formal errors in the fitted parameters of the relation. 

We obtained an alternative mass estimate by measuring the velocity dispersion and assuming dynamical equilibrium.
Direct application of the virial theorem \citep[][and refs. therein]{girardi1998} yields a value of $\sim 4.1\pm 2.7$ \mdod for the cluster mass contained within the sampled region, that is, $R \sim 0.26$ Mpc, which corresponds to the distance of the most distant galaxy from BGG1.
Using equation~1 from \citet{munari2013}, we obtain $M_{200,\sigma}=3.9\pm2.8$ \mdod within $R_{200,\sigma}=0.31\pm0.07$ Mpc. In this case, the large uncertainty in the mass value results from the large percentage error in velocity dispersion, since $M_{200}\propto \sigma_v^3$, with an additional 10\% due to the scatter in the \citet{munari2013} relation. 
Although the small number of group members and the resulting large uncertainties prevent a precise conclusion, the nominal difference between the two virial estimates and the one based on optical luminosity above is due to the observed velocity dispersion $\sigma_V\sim 160$ \kss. In fact, this is less than the virial value obtained from 
$M_{200,L}$, $\sigma_V\sim 270$ \kss, using the equation of \citet{munari2013} applied in reverse.
We suggest that this discrepancy between the virial and observed velocity dispersions arises because galaxies can be slowed by 
two-body interactions, which are expected in group environments where the velocity dispersion of galaxies is comparable to that of stars in luminous galaxies. This could suggest that this group is in a collapsing phase.

\subsection{BGG1 and the dynamical status of the \emi group}
\label{dyn}

Since this environment likely favors multiple interactions and collisions among galaxies, it is worth further analyzing and discussing BGG1 and the group's dynamical status. Figure~\ref{compact} shows a zoomed-in image around BGG1 in the top right of the panel. There are at least three faint objects on the northwest side, very close to  BGG1. The most interesting is the northern object, which is extended, has low surface brightness, and is entirely embedded in the light envelope of BGG1. The Legacy Survey DR10 lists it as a  round-type object with a $g$-$r$ color similar  to that of BGG1. 
The optical spectrum of BGG1 shows several emission lines. The most prominent are [OII] and [NII]. High-ionization emission lines are not present, and we classify it as a low excitation galaxy (LEG).

Since the five galaxies in the central part of the \emi group are especially concentrated in a dense core, we also considered the possibility that these five galaxies form a compact group (IDs 1-5 in Table~\ref{tabgroup}, see Fig.~\ref{compact}).  The photometric criteria of \citet{hickson1982} define a group as compact if two circles with radii $\Theta_{G}$ and $\Theta_{N}\ge 3~\Theta_{G}$ can be determined such that: 1) within the radius $\Theta_{G}$, there are $N \ge 4$ galaxies within a range of $3~\ma$ from the brightest galaxy; 2) between $\Theta_{G}$ and $\Theta_{N}$, there is no other galaxy in the above magnitude range or brighter (the isolation criterion); 3) the surface brightness within $\Theta_{G}$
is $\mu_{\rm G}< 26.0~\maa$. Figure~\ref{compact} shows the circle containing the five galaxies closest to BGG1 and a circle three times larger. The five galaxies are BGG1, BGG2, and three others, all within the required magnitude range, yielding $\mu_{\rm  G}=22.8~\maa$. A brightest value of $\mu_{\rm  G}$ indicates the compact-group nature according to \citet{mendel2011}.
Figure~\ref{compact} shows the isolation region included between $\Theta_{G}$ and $\Theta_{N}$. All objects within the isolation region are stars or faint galaxies, except for  galaxy ID~6 (see Fig.~\ref{compact}).
This galaxy is 2.84 magnitudes fainter than BGG1 in the $g$ band \citep[as suggested by][]{hickson1982} and 2.96 magnitudes fainter in the $r$ band. Thus, the isolation criterion is not perfectly met and we define the core of the \emi group as a quasi-compact group.

The core of the \emi group can be compared with compact groups  identified in mock redshift-space galaxy catalogs by \citet{diaz-gimenez2021}. The most robust observational parameter is the magnitude difference between BGG2 and BGG1, $M_2-M_1=1.01$. Comparison of this value with the corresponding panel in Figure~9 of  \citet{diaz-gimenez2021} suggests that this is a compact group that did not form early, but rather during a later phase (via late, secondary, or gradual formation channels). Although a spurious compact group cannot be excluded on this basis alone, the low velocity dispersion, $\sigma_{v,{\rm CG}}=162$ \kss, and the ratio of mass-to-light ratio, $M_{\rm vir}/L=43 h M_\odot/L\odot$ (in their units) suggest that a spurious group can be excluded.

\subsection{A786 and large-scale structure}
\label{env}

The environment of the \emi group shows a matryoshka-like structure. The group is located in the outskirts of A786, and the cluster is embedded in a rich supercluster (see Sect.~\ref{intro}). We studied A786 and the surrounding large-scale structure, taking advantage of the redshifts from DESI-DR1. We selected galaxies within a radius of 5\arcmin ~around the center of A786. 
Since the central cluster region is dominated by two bright elliptical galaxies surrounded by a common intracluster light, we fixed the center at the peak of the ROSAT X-ray emission \citep[R.A.=$09^{\mathrm{h}}28^{\mathrm{m}}04.94\dotsec$,
Dec.=$+74\degree46\arcmm 58.0\arcs$ - J2000.0,][]{boehringer2000}. In a sample of 33 galaxies, we detect a peak at $z\sim 0.123$, which confirms the previous cluster redshift estimate.

\begin{figure}
\centering
\resizebox{\hsize}{!}{\includegraphics{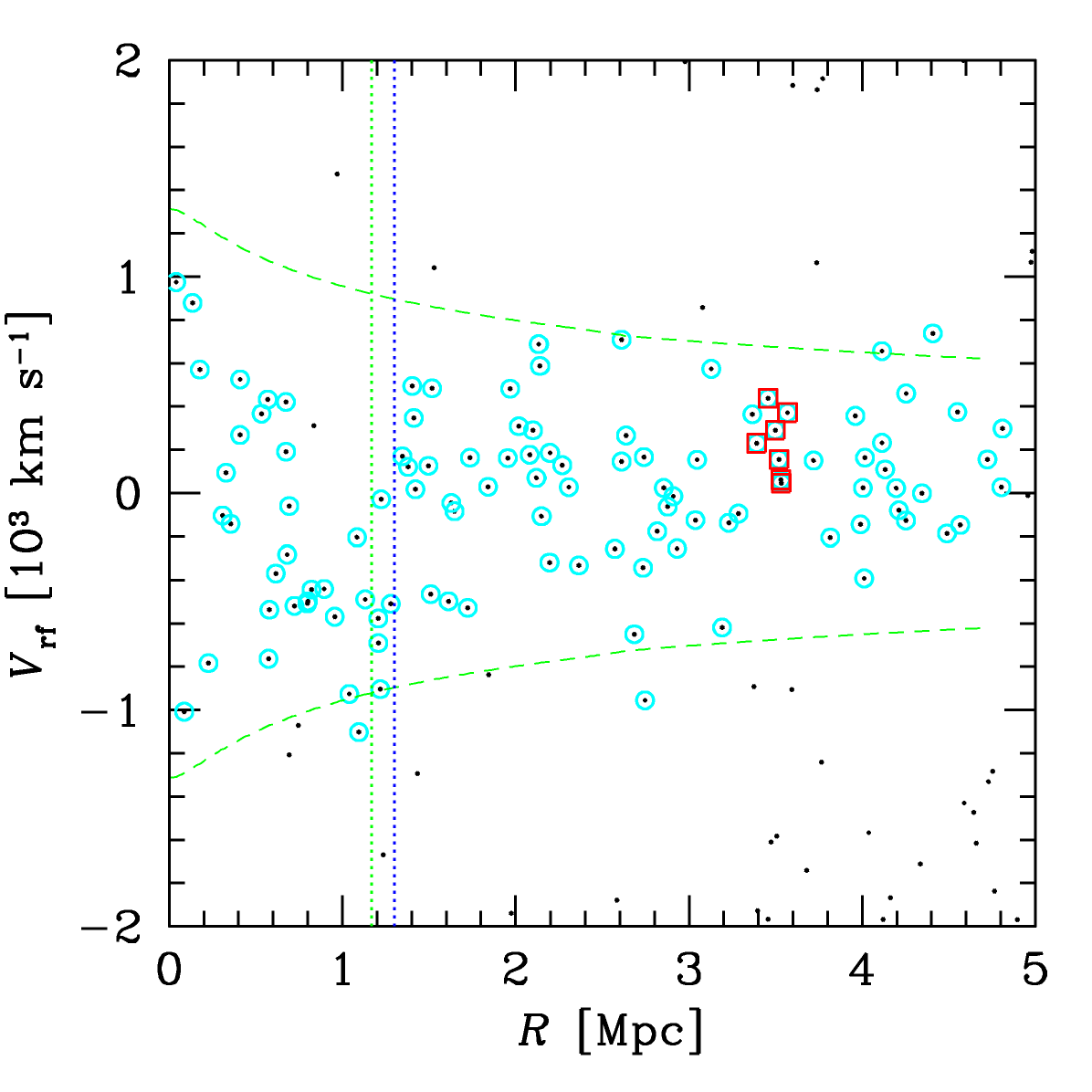}}
\caption
      {Projected phase space of the galaxies belonging to the densest redshift peak in the A786 region.  The rest-frame velocity,
      $V_{\rm rf}=(V-<V>)/(1+z)$, is shown as a function of the projected  distance $R$ from the A786 center (here, the X-ray center). Cyan symbols indicate member galaxies. The dashed green curves enclose the region where $V_{\rm rf}$ is lower than the      escape velocity (see text).  The caustics are shown out to
      $4\times R_{200}$. The vertical dotted green line marks the $R_{200}$ radius. The vertical blue dotted line marks an
      alternative $R_{200}$ estimate derived from the X-ray
      luminosity.  The red squares indicate the seven galaxies selected as members of the \emi group (see Sect.~\ref{gru}), located approximately $3\times  R_{200}$ from the A786 center.}
\label{figvd}
\end{figure}

To select the member galaxies of A786, we first examined the projected phase space and included galaxies within approximately 5~Mpc of the cluster center (specifically 37\arcmm; see Fig.~\ref{figvd}). This resulted in a sample of 1\,231 galaxies. We then applied the full procedure of member selection known as ``peak + gap'' \citep[P+G;][]{girardi2015}. We used the 1D-DEDICA algorithm to identify the cluster redshift peak of 156 galaxies at $z\sim 0.123$. In the second step of the procedure, we combined galaxy positions and velocity information. We worked in projected phase space, where galaxies belonging to galaxy systems appear within regions bounded by caustics with a characteristic trumpet shape. We used the ``shifting gapper'' method \citep{fadda1996,girardi1996}. For galaxies located within an annulus around the center of the system, the method excludes those that are too far away in velocity from the main body of galaxies, that is, those farther than a specified velocity gap. The annulus is shifted outward with an increasing distance from the center of the cluster.
The procedure is repeated until the number of cluster members converges to a stable value.  Following \citet{fadda1996}, we used an annulus size of 0.6 Mpc or larger to include at least 15 galaxies.  For the velocity gap, we adopted a value of
500 \kss, which has been used for very well-sampled systems or small galaxy systems \citep{girardi2023, girardi2024}.  Among the 156 galaxies in the redshift peak, we selected 111 members for A786.  The member galaxies are in the $35\,682<cz<38\,015$ \ks range.

For A786, we used the 111 members to calculate the mean line-of-sight velocity $\left<{V}\right>=36\,920\pm$41 \kss ($z=0.1232\pm0.0001$) and the velocity dispersion $\sigma_{V}=433_{-31}^{+32}$ \kss. Assuming that A786 is virialized, we derived the mass $M_{200}$ from the velocity dispersion by applying Eq.~1 of \citet{munari2013} and using the recursive approach as in \citet{girardi2024}. We obtain $M_{200}=(2.0\pm0.9)$ \mquaa. We used the 29 galaxies within $R_{200}=(1.2\pm 0.1)$ Mpc to compute $\sigma_{V,200}=580_{-57}^{+80}$ \kss. We calculated the uncertainty for $R_{200}$ using error propagation for $\sigma_{V}$ ($R_{200}\propto \sigma_{V}$), and we estimated the uncertainty for $M_{200}$ using error propagation ($M_{200}\propto \sigma_{V}^3$), with an additional 10\% uncertainty to account for the scatter in the relation from \citet{munari2013}. Alternatively, using the X-ray luminosity of \citet{boehringer2000} and the $L_X-M_{200}$ relation of \citet{reiprich2002}, we obtain a mass $M_{200, Lx}\sim 3$ \mqua, in agreement with the above estimate.

Figure~\ref{figvd} summarizes the above analysis and shows the projected phase-space distribution of the galaxies in A786.  The escape-velocity curves, calculated following \cite{denhartog1996}, are also shown. In the computation, we assumed a Navarro, Frenk and White (NFW) mass density profile \citep[][]{navarro1997}, adopted the concentration parameter from the relation of \citet{dolag2004}, and applied the mass estimate obtained above from galaxy kinematics. This serves as a posteriori verification of our membership procedure and shows that the galaxies of the \emi group lie within the caustic region of A786. at a distance of about $3\times R_{200}$ from the center of A786.
Other examples of compact groups in the outskirts of galaxy clusters have been reported in the literature \citep[e.g.,][]{pompei2006,pompei2012,jaffe2013}.

\begin{figure}
\centering
\resizebox{\hsize}{!}{\includegraphics{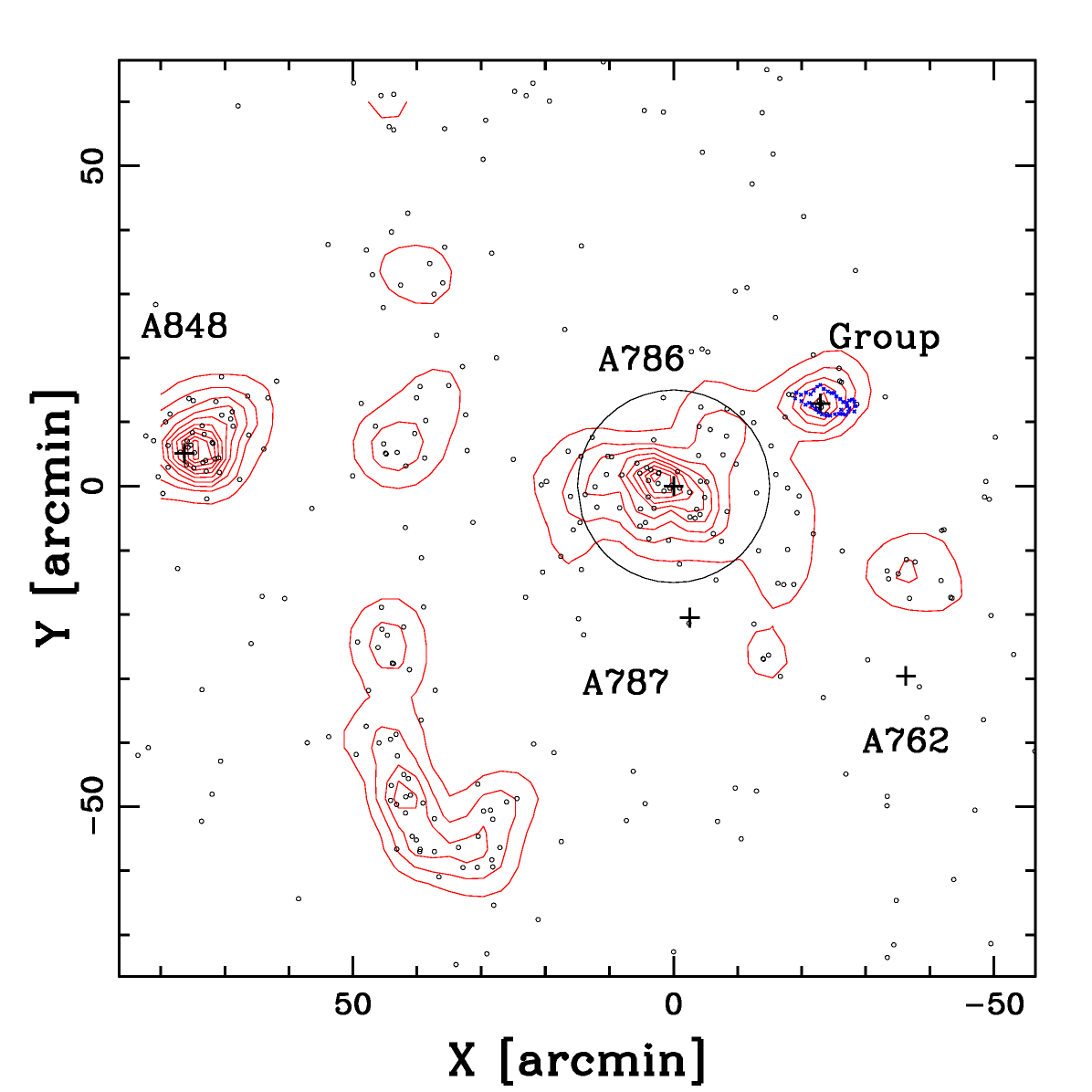}}
\caption
    {Spatial distribution and 2D-DEDICA isodensity contours of the      galaxies in the redshift range of A786 ($35\,682<cz<38\,015$). The circle highlights the region within a radius of 15\arcm ($\sim 2$ Mpc) around the cluster center, marked by a cross. 
    An second cross marks the \emi group (labeled as `Group''). A786 lies  
      at the western edge of the supercluster SCL245, which also includes A848 \citep{chowmartinez2014}. 
      The positions of A762 and A787 are also indicated. Although they are close to the \emi group in sky position , they belong to a background supercluster and are therefore not detected in the A786 redshift range. The blue contour marks the position of the radio source.
      }
\label{figk2}
\end{figure}

To better understand the connection between the \emi group and its surroundings, we considered the sample of 496 galaxies located within two degrees of BGG1 and with $35\,682<cz<38\,015$, placing them in the same redshift range as the A786 member galaxies. We performed a 2D-DEDICA analysis to detect galaxy concentrations on the sky. Figure~\ref{figk2} presents our results in the region with notable structures. The main 2D peaks in the  analyzed region are  A786 and A848, two Abell clusters that both belong to the supercluster SCL245, with A786 at its western edge  \citep[]{chowmartinez2014}. We detect the \emi group with a confidence level greater than $99.99\%$. To its southeast, the \emi group appears connected to A786, while there is no evidence of a structure to its northwest. In this figure, we also show the positions of A762 and A785, two Abell clusters that are close to A786 in sky position but belong to the supercluster SCL203. SCL203 lies at a slightly higher redshift than SCL245 \citep[]{chowmartinez2014}; consequently, A762 and A787 are not detected when selecting galaxies in the A786 redshift range.

\section{Radio data}
\label{radio}

\subsection{LOFAR data}

The source was observed on 11 November 2014 with the LOFAR \citep[]{vanhaarlem2013} high band antenna (HBA) at 144 MHz in the HBA\_DUAL\_INNER mode, for a total observing time of 7.3 h (Project LC2\_008).
We calibrated the data using the standard procedure by \cite{shimwell2022}. Briefly, the data were first corrected for direction-independent (DI) effects \citep[for more details see][]{vanweeren2016calib, williams2016, degasperin2021} and then for direction-dependent (DD) effects using a pipeline based on the software tools killMS and DDFacet \citep{tasse2014a, tasse2014b, smirnov2015, tasse2018}. Finally, to improve the calibration in the target region, we followed the extraction and self-calibration procedure described by \cite{vanweeren2021}. We set the flux-density scale according to \cite{scaife2012heald} and the flux calibration uncertainty is estimated to be around $10\%$  \citep{Hardcastle2021, shimwell2022}. We then imaged the calibrated data at different resolutions using WSclean \citep{offringa2014}.

\begin{figure}
    \centering
    \includegraphics[width=\linewidth]{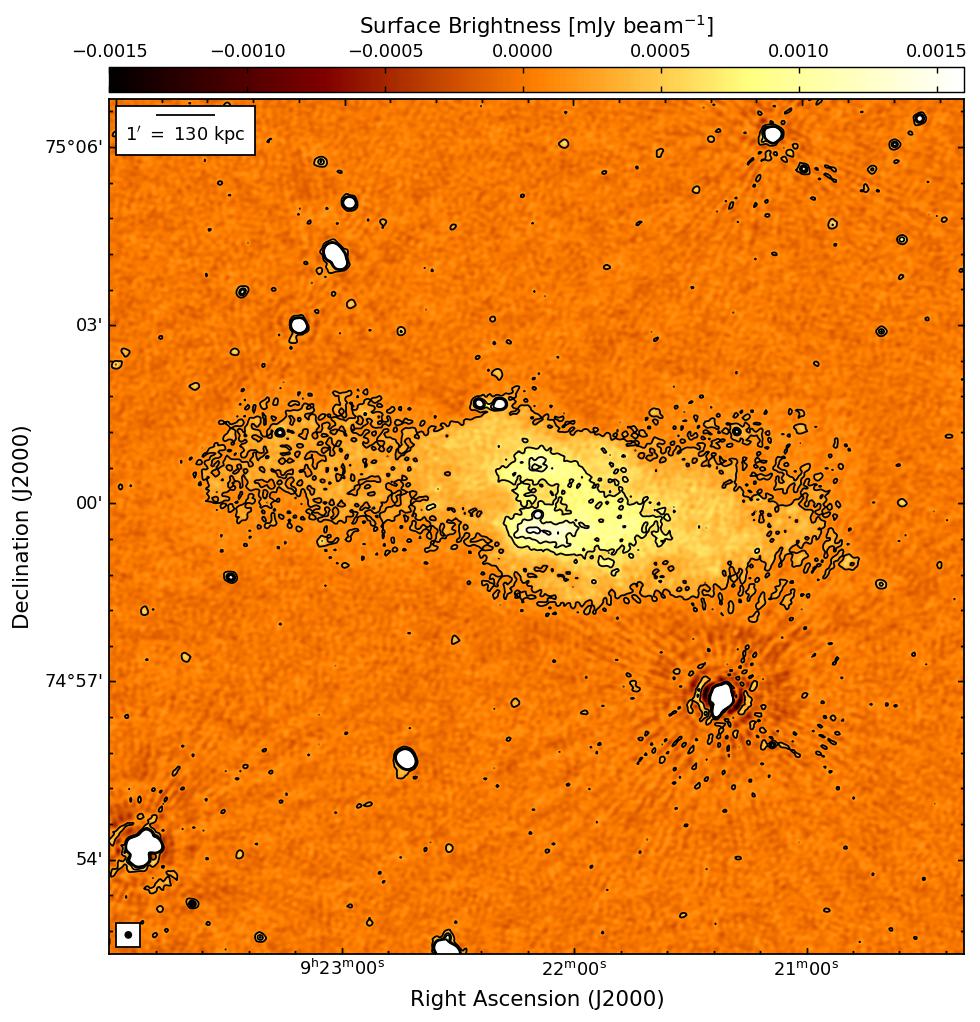}
    \caption{LOFAR image at 144 MHz with HPBW of 6 arcsec. Contour levels are [3, 10, 15, 20]$\times\sigma$, where the noise level is $\sigma=73\ \rm{\mu Jy/beam}$.}
    \label{Lofar}
\end{figure}

Figures \ref{Lofar} and \ref{compact} present the images obtained at 6 and 20 arcsecond resolution, respectively. Despite the different resolutions, the images show a similar source extent, demonstrating that no additional low-brightness emission is present. The high-resolution image indicates the lack of any structure in the large-size emission and the uniform brightness of this source. In the inner region, both images show a brighter region with some substructure, but no central core-jet active region is evident. 
In the high-resolution image, discrete emission identified with the BGG1 galaxy is visible 
(see Sect.~\ref{SecMor}). 
This source is not visible in the low-resolution image because of the presence of a substructure resembling a tailed source. This is a convolution artifact; the BGG1 radio emission is not the head of the tailed emission.

\subsection{VLA data}
We obtained observations in the L band by combining data from the projects AJ283, AG823, and AG344 using the Astronomical Image Processing System (AIPS). In these projects, the source \emi was observed in the C and D configurations in the L band. At 4.7 GHz, we used the D-configuration observation from project AJ283. 

We calibrated the datasets separately in AIPS following the suggestions in the cookbook. In all datasets, we used 3C 286 as the main gain calibrator. We also calibrated the polarized intensity (Q and U) and used 3C 286 to calibrate the polarization vector position angle.
After calibration, we used  the AIPS task DBCON to combine the L-band datasets.
We edited the calibrated datasets at both frequencies using TVFLG,  and performed phase self-calibration cycles to refine antenna phase solutions. Finally, we performed a final amplitude and phase cycle to remove residual gain variations. We applied a similar procedure to the single dataset at 4.7 GHz.

We used these data to obtain images using the full uv range and images at matched uv coverage to compare source properties at different frequencies without a coverage bias. 
The calibration errors on the measured flux densities are estimated to be $\sim$ 3 percent. All images have been corrected for primary-beam attenuation. 

\begin{figure}
\centering
\resizebox{\hsize}{!}{\includegraphics{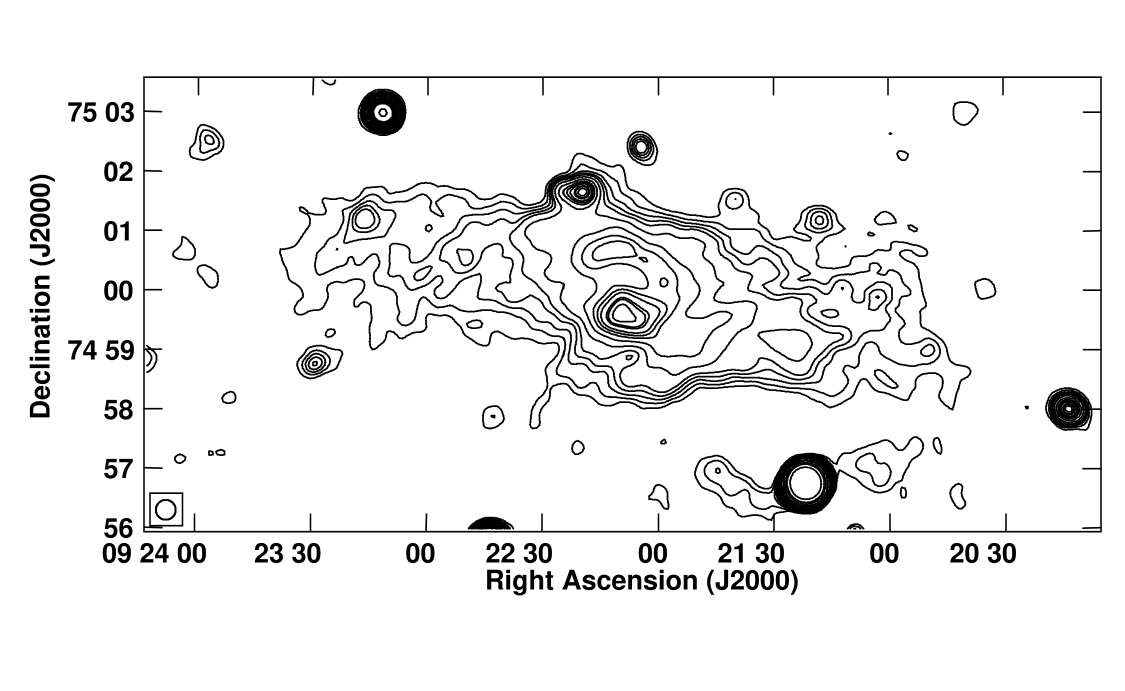}}
\caption
{Image of the source \emi at 1.4 GHz with a circular HPBW of 20 arcsec. The noise level is 47 microJy/beam. Contour levels are 0.1, 0.2, 0.3, 0.4, 0.5, 0.7, 1, 1.3, 1.5, 1.7, 2, 2.1, 2.3, 2.5, 3, 5, 6, 10, 15, and 30  mJy/beam. }
\label{L-full20}
\end{figure}

In Figure~\ref{L-full20}, we show the L-band image of the extended source with an HPBW = 20 arcsec obtained with the AIPS task IMAGR. A large field was reduced to clean the contribution of background sources. To properly image the diffuse extended emission we used a low clean factor.
We note that the presence of a strong source near the diffuse emission 
 (a few arcminutes southwest of \emi) introduces artifacts in the brightness distribution. The noise level is dominated by the dynamic range (1/500). Levels are reported in millijansky.

\begin{figure}
    \centering
\resizebox{\hsize}{!}{\includegraphics{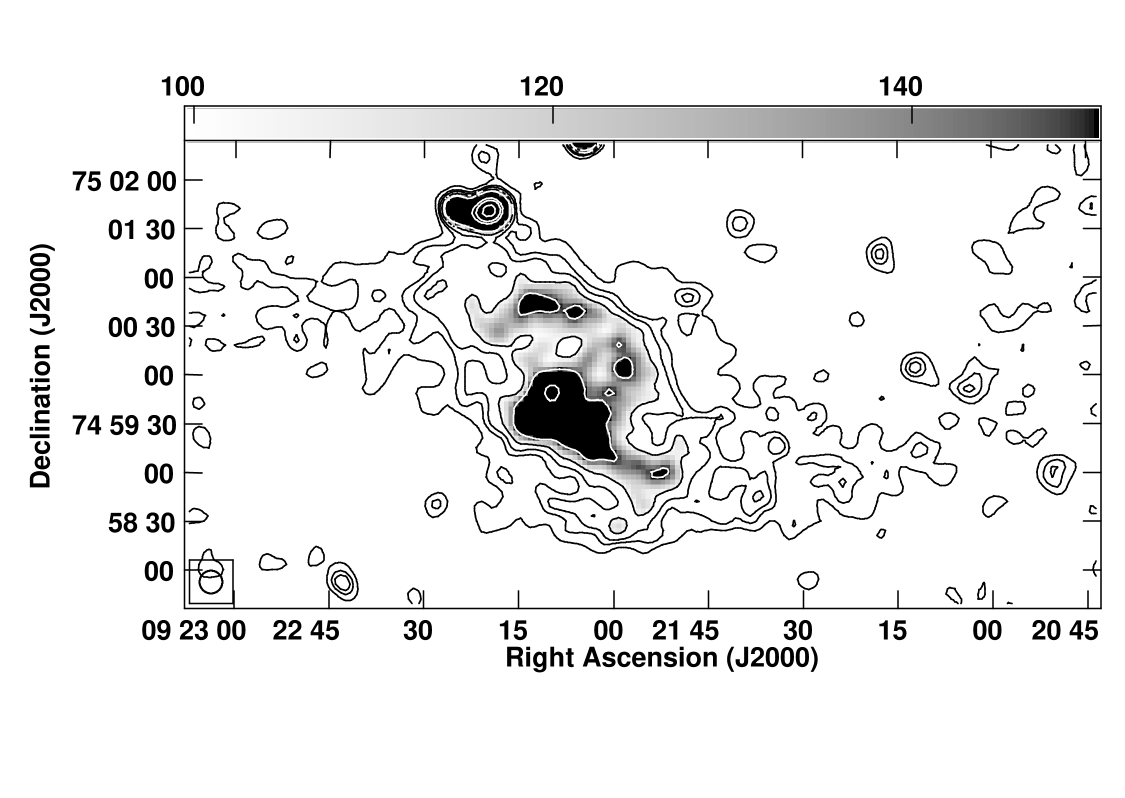}}
\caption
{Image of the source \emi at 4.7 GHz with a circular HPBW of 14 arcsec. The noise level is 15 microJy/beam. Contour levels are 0.02, 0.05, 0.07, 0.1, 0.15, 0.4, 0.5, 0.7, 1, 5, and 10  mJy/beam. The gray-scale range is 100 - 150 microJy/beam.}
\label{C14-gray}
\end{figure}

Figure~\ref{C14-gray} presents the C-band image with HPBW = 14 arcsec. 
In this image only the central inner region of \emi is visible due to the lack of short baselines at 1.4 GHz and the primary-beam attenuation, which affects such an extended source at 4.7 GHz. At this frequency, the VLA primary-beam cutoff is 9.13 arcmin (see AIPS task PBCOR), but the noise level already increases significantly at $\sim$5 arcmin from the image center.

\subsection{Source properties}

\subsubsection{Source parameters and morphology}
\label{SecMor}

Table~\ref{observational} lists the full source parameters.

\begin{table*}[ht]
    \centering
    \caption{Observational parameters of \emi.}
    \begin{tabular}{cccccc}
        \hline
        \noalign{\smallskip}
        Frequency & uv-range & HPBW  & Noise  & Total Flux & Size   \\
           GHz    & k-lambda & arcsec& $\rm{\mu Jy/beam}$& mJy        & arcmin \\    
        \hline
        \noalign{\smallskip}
0.15  & full       & 20 &144 & 784.59 & 11.3 x 3.8 \\
0.15  &0.38 - 16.55& 14 & \phantom{1}73 & 708.89 & 10.3 x 3.1 \\
1.43  & full       & 20 & \phantom{1}47 & 106.44 & 11.0 x 3.5 \\
1.43  &0.38 - 16.55& 14 & \phantom{1}30 & \phantom{1}92.11  & \phantom{1}9.8 x 3.1 \\
4.71  & full       & 20 & \phantom{1}21 & \phantom{1}13.70  & \phantom{1}6.7 x 3.2 \\
4.71  &0.38 - 16.55& 14 & \phantom{1}15 & \phantom{1}12.66  & \phantom{1}5.2 x 3.3  \\
\noalign{\smallskip}
        \hline 
        \end{tabular}
        
\tablefoot{Col.~1: Observing frequency. Col.~2: u-v range of different images. Col.~3: Angular resolution (HPBW). Col.~4: Noise level. Col.~5: Total flux density. Col.~6: Angular size}

    \label{observational}
\end{table*} 

\begin{table}[ht]
    \centering
    \caption{BGG1}
    \begin{tabular}{cccc}
        \hline
        \noalign{\smallskip}
        Frequency & R.A.,\   Dec.    & S       & notes   \\
           GHz    & h m s,\ °  $^\prime$ $^{\prime\prime}$ & mJy     &          \\    
        \hline
        \noalign{\smallskip}
0.15  & 09 22 09.55,+74 59 47.3 & 2.56 & HPBW 6" \\
4.7   & 09 22 09.73,+74 59 48.5 & 0.43 & HPBW 14" \\
optical&09 22 09.57,+74 59 48.5 & --   &  \\
\noalign{\smallskip}
        \hline
   \end{tabular}
   \tablefoot{Col.~1: Observing frequencies, or ``optical'' for optical data. Col.~2: R.A. and Dec. position of radio peaks and optical galaxy. Col.~3: Flux density (radio only). Col.~4: Radio angular resolution.}
    \label{BCG1}
\end{table} 

\begin{table}[ht]
    \centering
    \caption{Physical properties of the sources}
    \begin{tabular}{ccccc}
        \hline
        \noalign{\smallskip}
        Name & Frequency & LS   & Log P & M$_r$   \\
             &    GHz    & kpc   & W/Hz  &         \\    
        \hline
        \noalign{\smallskip}
\emi  &  0.15  & 1505 & 25.49  &  -- \\
\emi  &  1.43  & 1465 & 24.63  &  -- \\
BGG1  &  4.71  &$<$ 20& 22.24  & -22.4 \\
\noalign{\smallskip}
        \hline
   \end{tabular}
   \tablefoot{Col.~1: Source name, including the full source and the discrete source identified with BGG1. Col.~2: Observing frequency. Col.~3: Linear Size. Col.~4: Log of the radio power. Col.~5: Absolute $r$-band magnitude of BGG1.}
    \label{physical}
\end{table} 

\begin{figure}
    \centering
\resizebox{\hsize}{!}{\includegraphics{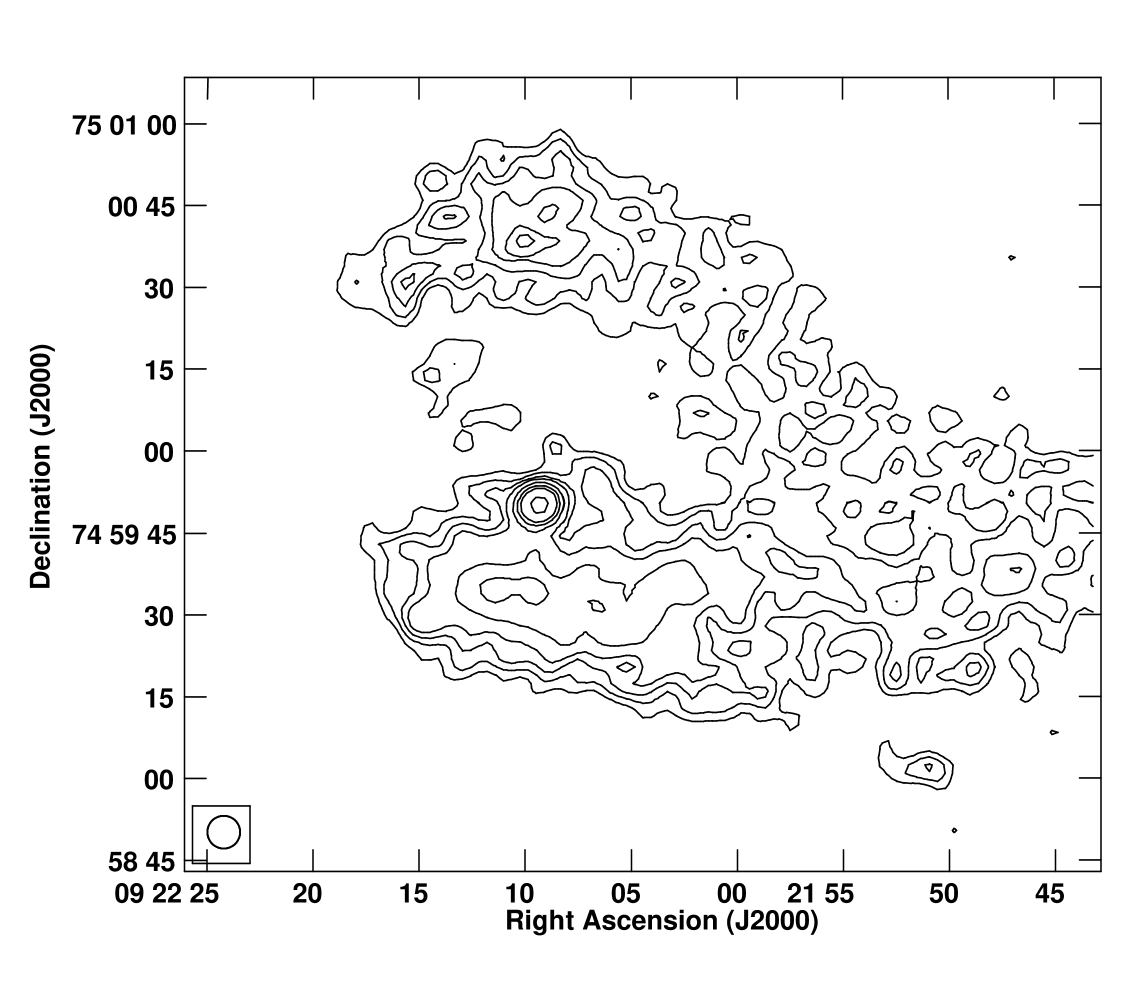}}
\resizebox{\hsize}{!}{\includegraphics{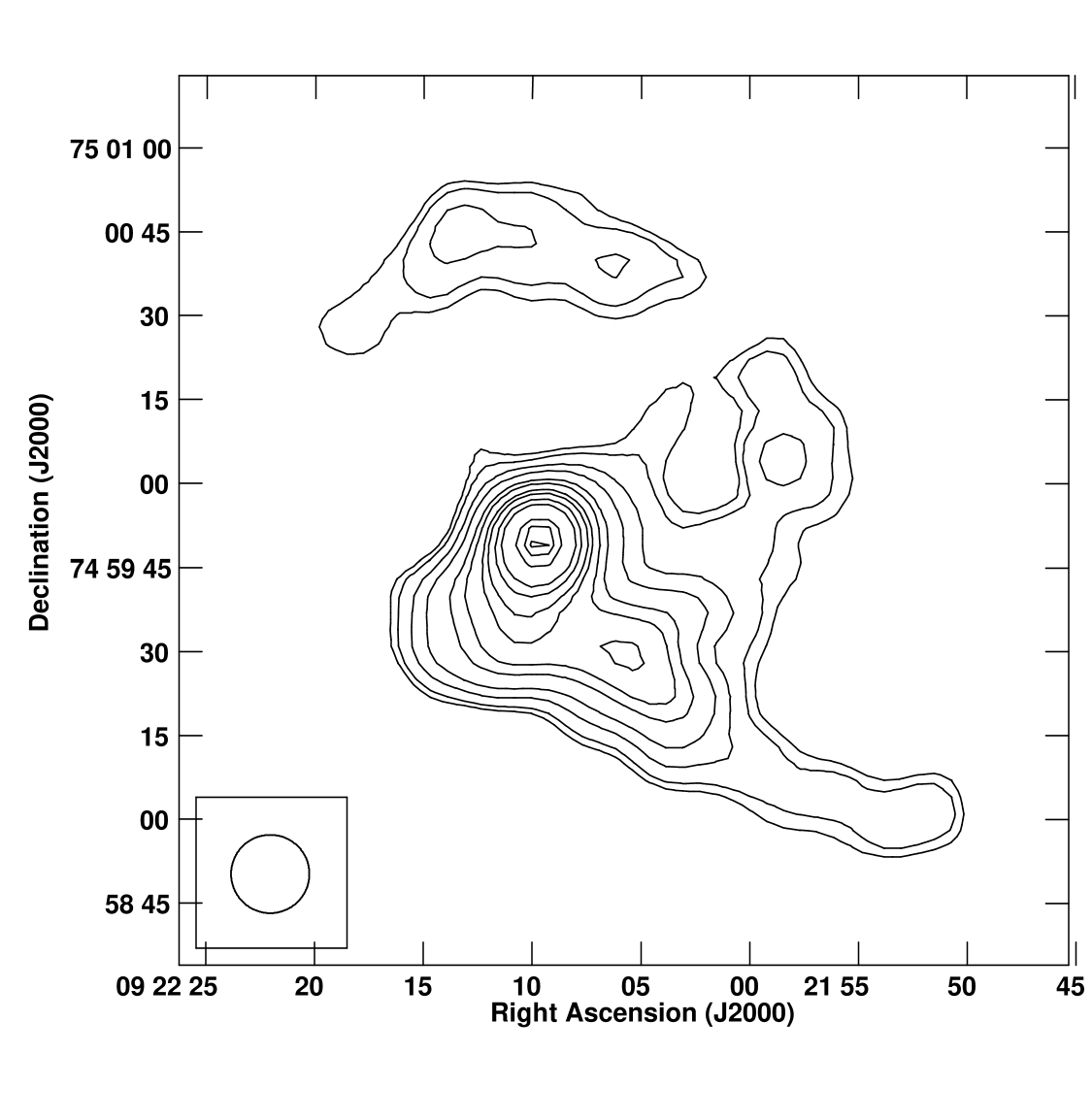}}
\caption
{Top: LOFAR 144 MHz image of the central region of \emi with a circular HPBW of 6 arcsec. The noise level is 80 microJy/beam. Contour levels are  0.8, 0.9, 1, 1.1, 1.3, 1.5, 1.7, and 2.3 mJy/beam.
Bottom: Image at 4.7 GHz of the inner region of \emi with a circular HPBW of 14 arcsec. The noise level is 21 microJy/beam. Contour levels are 0.12, 0.13, 0.15, 0.17, 0.2, 0.22, 0.25, 0.27, 0.3, 0.33, 0.4, 0.43, and 0.45 mJy/beam. In both images, levels have a high signal-to-noise ratio to better highlight the central inner structure, which is superimposed on the extended diffuse emission.}
\label{core}
\end{figure}

This giant radio source is elongated in the east-west direction (PA $\sim$ 80$^\circ$), with a ratio of about three with respect to the north-south extension. The external regions imaged with LOFAR and in the L band are resolved with a diffuse surface brightness and show no hotspots, which is typical of Fanaroff-Riley Class I (FR I) radio galaxies, but without any jet-like structure. The west region is slightly brighter than the east region. The few discrete emission regions are clearly sources unrelated to \emi, with the exception of the discrete source coincident with the BGG1 galaxy (see Fig.~\ref{L-full20} and Table~\ref{BCG1}).  
The central region shows a slight S-shaped structure.

Figure~\ref{core} shows the central region of \emi in the LOFAR data and at 4.7 GHz, allowing a detailed comparison of the structures present in the central region. At the highest resolutions, a spiral-like structure is visible emanating from the point-like source and pointing south with two distinct arms. The upper arm curves toward the east in a circular shape, while the southern arm extends toward the west. This spiral-like structure covers an area of about 2 x 2 arcmin (250 x 250 kpc). The separation between BGG1 and BGG2 is 1.6 arcmin (about 200 kpc; see Table~\ref{tabgroup} and Fig.~\ref{compact}). This structure is surrounded by a relatively diffuse brighter region at PA $\sim$ 40$^\circ$, with a linear size (LS) of $\sim$ 440 kpc (clearly visible in Fig.~\ref{C14-gray}). This region slowly merges into the giant east-west structure discussed above (Figures~\ref{compact} and \ref{L-full20}).  

Table~\ref{physical} lists the physical properties of \emi and BGG1, assuming that \emi lies at the same distance from BGG1. The LS ($\sim$ 1.5 Mpc) and the total radio power are consistent with giant radio sources \citep[see e.g.,][]{Dabhade2020}. We note that the BGG1 radio parameters are difficult to estimate because BGG1 is embedded in the \emi emission. We used a Gaussian fit to the discrete source after subtraction of the extended emission.

To further study this region, we extracted radio images from the public available Very Large Array Sky Survey (VLASS)\footnote{for more details see  https://science.nrao.edu/vlass}. The data cover the frequency range of 2 to 4 GHz with an angular resolution of $\sim$ 2.5 arcsec . We did not detect any radio emission at the BGG1 position at a level of 0.35 mJy/beam (3$\sigma$); this suggests that the source is resolved out.

\subsubsection{Spectral index, age, and magnetic field}

To compare source properties at different wavelengths (0.15--4.7 GHz) and derive the spectral index distribution, we obtained LOFAR and VLA images with the same uv-range (0.38 - 16.55 k$\lambda$).
The integrated spectral index of the full source is $\alpha\sim0.89$ between 0.15 and 1.4 GHz and $\alpha\sim1.67$ between 1.4 and 4.7 GHz (limited to the region visible at 4.7 GHz).

\begin{figure}
    \centering
    \includegraphics[width=\linewidth]{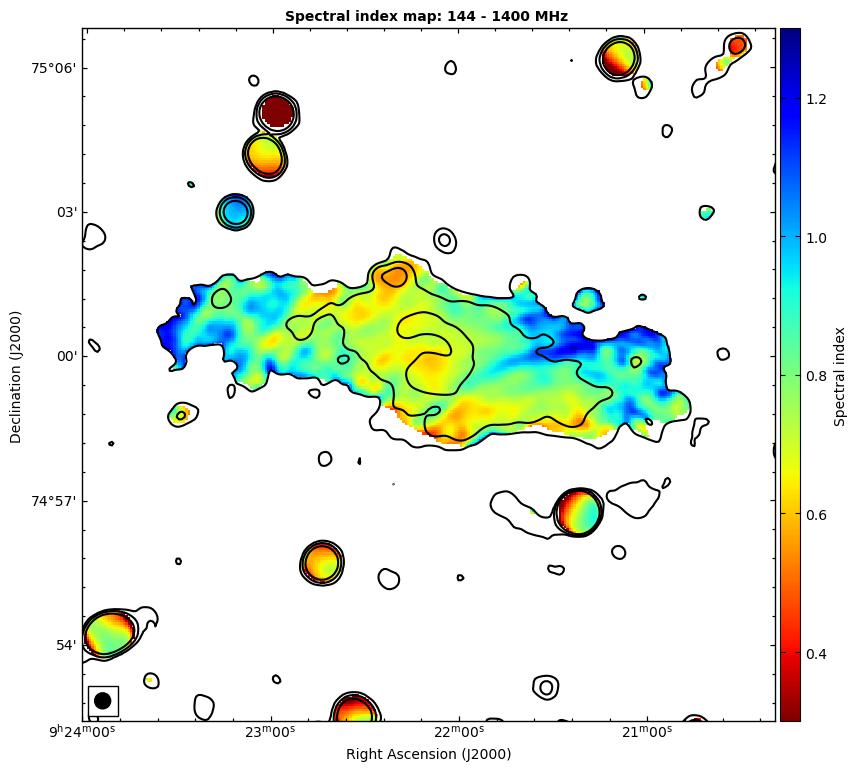}
    \caption{Spectral index map between 144 MHz and 1.4 GHz at 20 arcsec resolution, with overlaid VLA L-band contours. }
    \label{spix20}
\end{figure}

To investigate the spatial distribution of the spectral index across the source extension, we produced a two-frequency spectral index map at 20$\arcsec$ resolution using LOFAR 144 MHz and VLA 1.4 GHz images (Figure~\ref{spix20}). We note that the bright source located southwest of the target produces artifacts that affect the study of the diffuse emission.
The central region has a mean spectral index of $\alpha=0.70\pm0.05$, while the outer emission that extends east-west has slightly steeper spectra, reaching 
$\alpha=0.9\pm0.1$ and up to $\alpha = 1.1\pm0.2$ in most external regions. However, there is no sharp spectral-index variation that would indicate the presence of two different electron populations, as in restarted radio galaxies. 

\begin{figure}
    \centering
    \includegraphics[width=\linewidth]{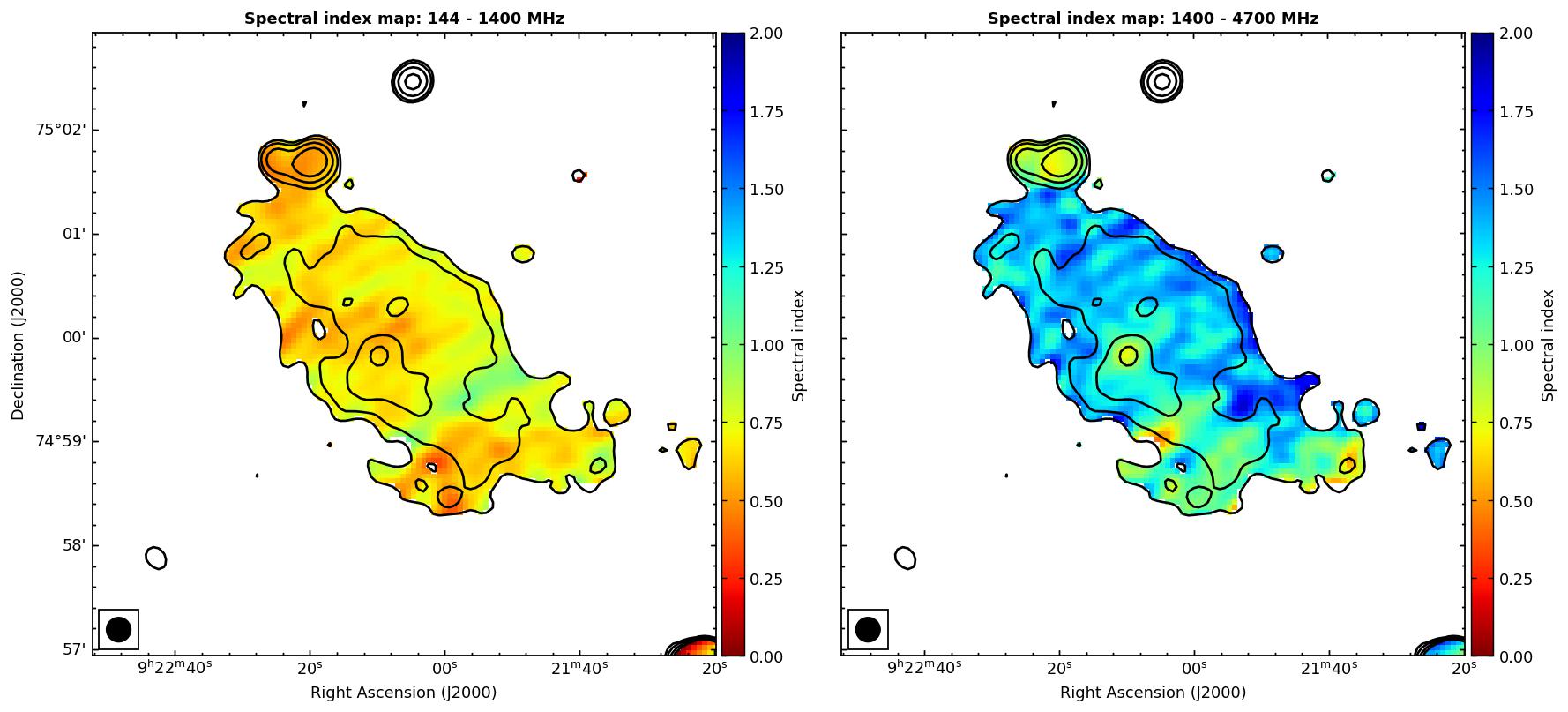}
    \caption{Spectral index maps between 144 MHz and 1.4 GHz (left) and between 1.4 GHz and 4.7 GHz (right) at 14 arcsec resolution, with overlaid VLA C-band contours. Minor bands present in images are artifacts of a strong nearby source; see Sect. 3.2}
    \label{spix_all}
\end{figure}

Including the 4.7 GHz data in the spectral analysis restricts the study to the central region of the emission at PA $\sim$ 40$^\circ$ (see Fig.~\ref{C14-gray}). In this case, we compared images obtained at an angular resolution of 14$\arcsec$ for a more detailed study of the source. 
We find a uniform steeper spectral index across the source extension in the higher-frequency range (1.4 - 4.7 GHz), consistent with the total spectral index in this region. This is shown in Figure~\ref{spix_all}, where we present spectral-index maps of the central emission of the source between 144 MHz and 1.4 GHz (left) and between 1.4 and 4.7 GHz (right), using the same color scale for direct comparison.

The spectral steepening observed at high frequency indicates the presence of a break in the spectra, implying the aging of cosmic rays. The break frequency relates to the time elapsed since particle injection and to the magnetic field strength as $\nu_b \propto B^{-3} t^{-2}$. Therefore, once the magnetic field is known, the age of the source can be derived from the shape of the spectra. To estimate the radiative age, we used the Broadband Radio Astronomy Tools (BRATS) code \citep{harwood2013}, which fits the observed radio spectrum with a model spectrum obtained by numerical integration of the equations that describe the radiative losses of the plasma. In particular we fit the spectra using the Jaffe-Perola (JP) \citep{jaffe1973} and Kardashev-Pacholczyk (KP) \citep{kardashev1962, pacholczyk1970} models of spectral aging, which account for synchrotron and inverse-Compton losses in a constant magnetic field.

To derive the magnetic field in the central region, we assumed equipartition between particles and the magnetic field. We find a value of $B_{eq}=0.63\ \rm{\mu G}$ with the classic formula and  $B_{eq}=0.75\ \rm{\mu G}$ with , while with the revised formula of \cite{beck2005}. \cite{chen2008} derived a lower limit of the magnetic-field strength from the non-detection of Inverse Compton emission in the source, finding $B\ge0.81\ \rm{\mu G}$. The low values of the derived magnetic field imply that the equivalent magnetic field strength of the CMB at the source redshift ($B_{CMB}=3.18(1+z)^2\ \rm{\mu G}=4 \rm{\mu G} )$ is the dominant factor in the age estimate.

\begin{figure*}
    \centering
    \includegraphics[width=\linewidth]{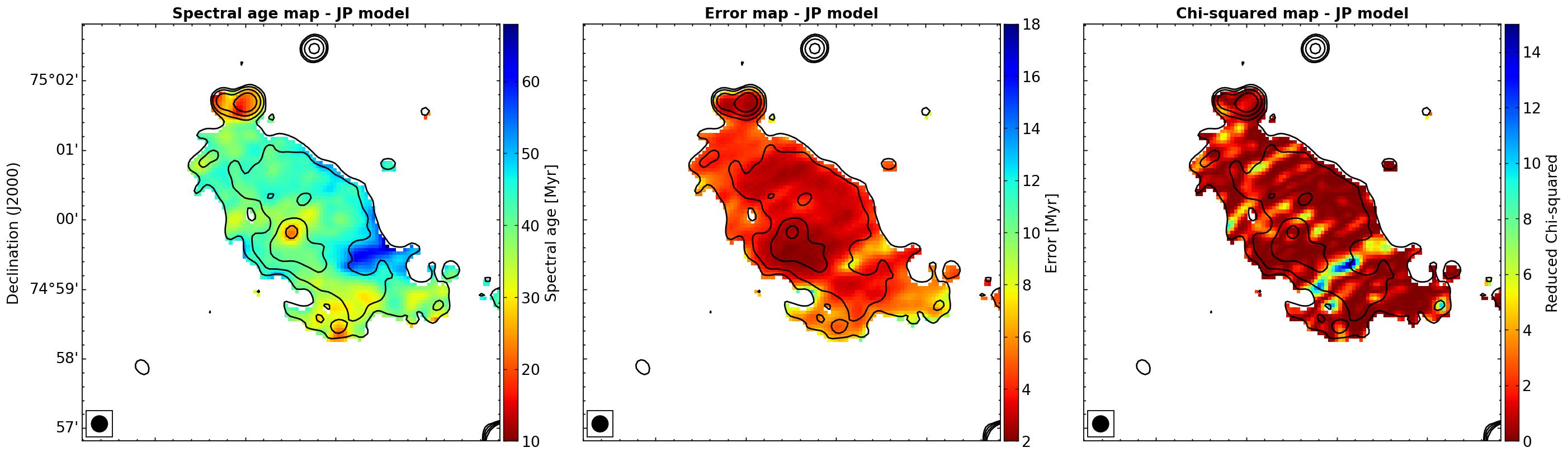}
    \caption{Jaffe–Perola (JP) spectral aging map (left), corresponding error map (middle), and reduced chi-squared map (right) of the central region of the radio source at 14 arcsec resolution. The C-band contours are overlaid. The fit assumes $\alpha_{inj} = 0.5$ and $B = 1.0\ \rm{\mu G}$.}
    \label{age_map}
\end{figure*}

We fit the models over a range of injection indices and find that the lowest chi-squared value is obtained with $\alpha_{inj}=0.5$ for both models. Fixing the injection index to 0.5, we derived the radiative age of the source using different values of magnetic field strength in the range $0.6\le B\le1.0$ $\rm{\mu G}$. We find similar age values for the JP and KP models.
Considering different values of magnetic-field strength and spectral aging models, and taking into account the spectral age errors, we find that the central discrete source identified with BGG1 has a radiative age in the range $t = 20 - 32$ Myr, while the surrounding diffuse emission has an age in the range $t=29-46$ Myr. 
Figure~\ref{age_map} shows the spectral age map with the corresponding error and chi-squared maps of the JP model, obtained using a magnetic field of $B=1\ \rm{\mu G}$. We find slightly shorter ages for lower magnetic-field values. We note that an age of 40 Myr implies an average expansion velocity of 0.02 c.

The large-scale diffuse emission extending in the east-west direction has steeper spectra, suggesting the presence of older particles. However, because it is detected at only two frequencies, we were unable to constrain its radiative age.

We can place constraints on the age of these large-scale regions on the basis of kinematical considerations (kinematic age).
Assuming BGG1 as the center of this source, the largest distance of the emission on both sides of \emi is 750 kpc, and assuming 
a reasonable expansion velocity of 0.02 - 0.05 c \citep[see before and e.g.,][]{Andernach1992}, the kinematic age of the external regions of the source is 50 - 100 Myr.

\subsubsection{Polarized data}

We obtained the Stokes parameter Q and U images at 1.4 and 4.7 GHz. Polarized data are not available from our LOFAR data. We obtained the final images with the same parameters as the total-intensity images and corrected the images for primary beam attenuation. We obtained the polarization intensity P = (U$^2$ + Q$^2$)$^{0.5}$ and the polarization angle PA = arctg(U/Q) from the Q and U images. The polarization-intensity images were corrected for the positive bias.

\begin{figure}
    \centering
\resizebox{\hsize}{!}{\includegraphics{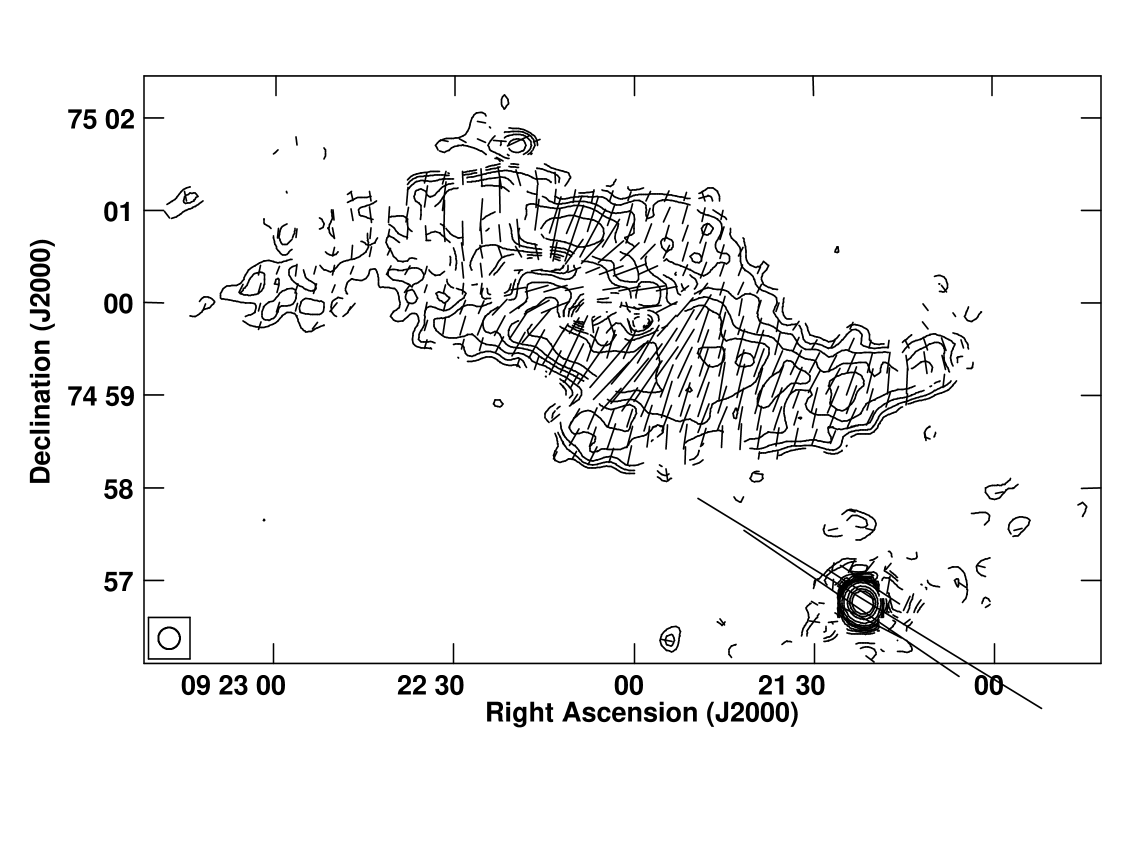}}
\caption
{Image of the polarized emission at 1.4 GHz with HPBW of 14 arcsec. Contours represent the intensity of the polarized emission and are 0.05, 0.07, 0.1, 0.15, 0.2, 0.3, 0.5, 1, 1.5, and 2 mJy/beam. Vectors are proportional in length to the total polarized intensity (POLC) with 50 arcsec corresponding to 0.56 mJy/beam and are oriented along the polarization angle (POLA).}
\label{POL-L1A}
\end{figure}

Figure~\ref{POL-L1A} shows the intensity of the polarized emission (contours and vector length) and relative PA (vector orientation) in the L band. Polarized emission is present along the whole source region, with the exception of the most external faint region, probably because of sensitivity limitations.

\begin{figure}
    \centering
\resizebox{\hsize}{!}{\includegraphics{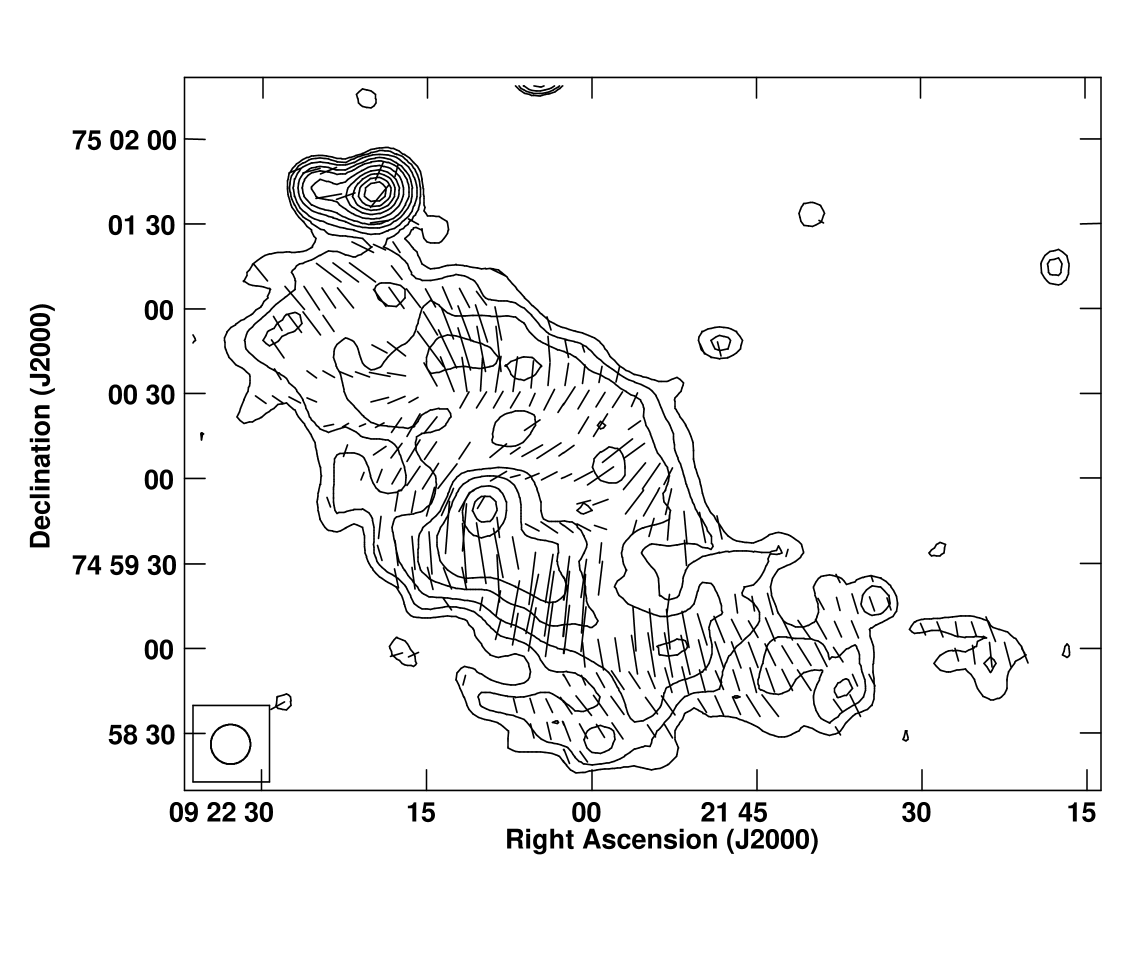}}
\resizebox{\hsize}{!}{\includegraphics{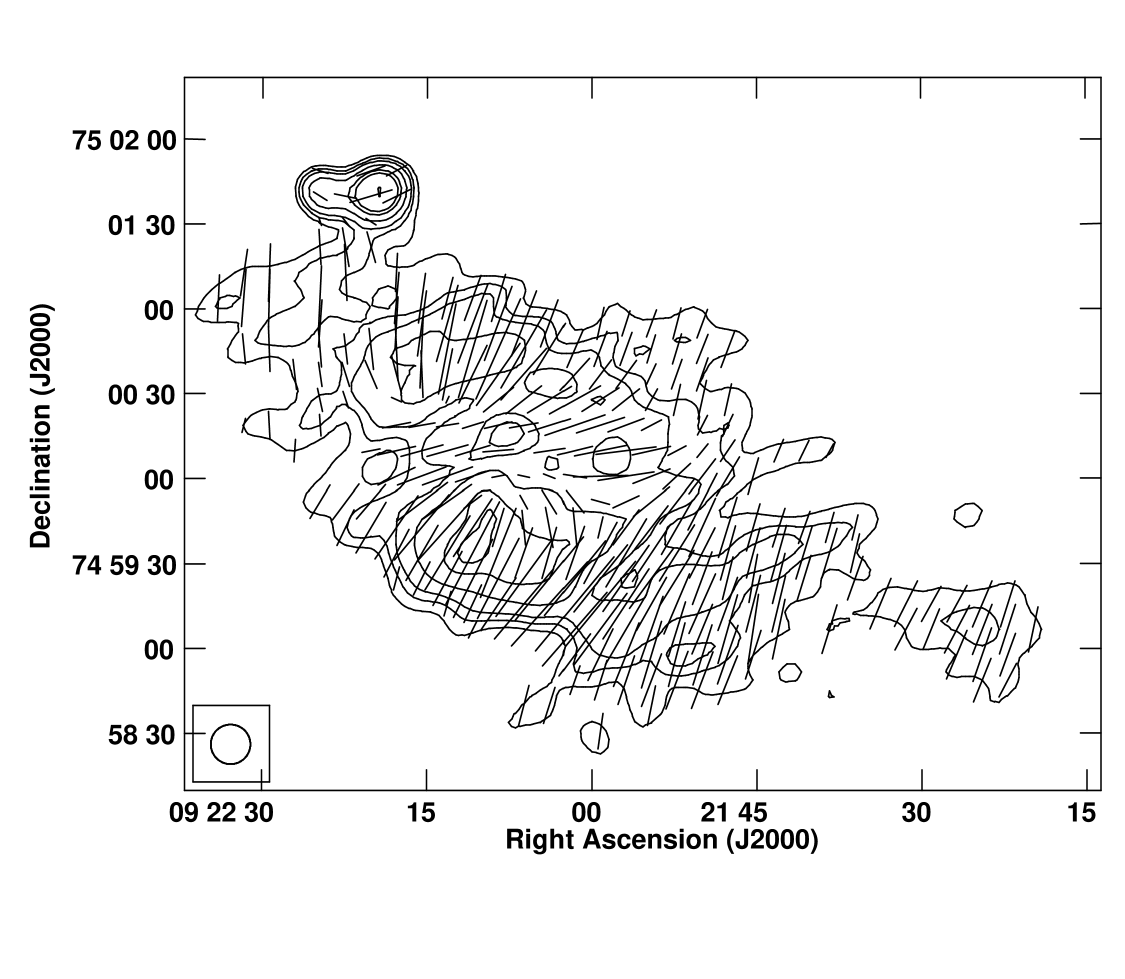}}
\caption
{Top: Total-intensity (I) of the source at 4.7 GHz with HPBW of 14 arcsec. Contour levels are 0.05, 0.07, 0.1, 0.15, 0.2, 0.3, 0.4, 0.5, 0.6, 0.7, and 1 mJy/beam. Superimposed vectors are proportional in length to the polarized intensity, with 20 arcsec corresponding to 0.11 mJy/beam, and are oriented as the polarization position angle. 
Bottom: Same as top panel, but at 1.4 GHz. Contour levels are 0.3, 0.4, 0.5, 0.7, 1, 1.2, 2, 3, 5, 10, 30, and 50 mJy/beam. A vector of 20 arcsec corresponds to 0.22 mJy/beam.  }
\label{POL-C1L1}
\end{figure}

At 4.7 GHz, only the central region of \emi is visible, as discussed in the previous section. 
In the top panel of Figure~\ref{POL-C1L1} we show the total-intensity emission at 4.7 GHz with the polarized emission shown as vectors whose length is proportional to the polarized intensity and whose orientation follows the polarization PA.

\begin{table}[ht]
    \centering
    \caption{Polarization results}
    \begin{tabular}{ccccc} 
        \hline
        \noalign{\smallskip}
region & Frequency &    I   &     P     & perc.  \\
       &    GHz    &   mJy  &    mJy    &      \%       \\    
        \hline
        \noalign{\smallskip}
total  & 1.43    & $92.1 \pm3\phantom{.5}$   &$32.5 \pm1$   &     35      \\
total  & 4.71    &$13.7 \pm0.5$ &$\phantom{00.}3.98 \pm0.09$ &     31      \\
Fig.\ref{POL-C1L1} & 1.43    &$77.3 \pm2\phantom{.5}$   &$\phantom{0.}29.3 \pm0.9$ &     38      \\
\noalign{\smallskip}
\hline
   \end{tabular}
   \tablefoot{Col.~1: Source region. Col.~2: Observing frequency. Col.~3: Total intensity flux density. Col.~4: Polarized intensity flux density. Col.~5: Polarization percentage.}
    \label{polar_parameters}
\end{table}

The bottom panel of Figure~\ref{POL-C1L1} shows the same field at 1.4 GHz, allowing a visual comparison of the polarization structure in the different source positions. At both frequencies, the polarized emission follows the total-intensity structure, including the spiral-like structure. The orientation of polarized vectors at the two frequencies is similar, implying a small rotation measure. This suggests that the magnetic field in the outer regions is approximately parallel to the source structure.  

The polarization percentage is higher at 1.4 GHz than at the other frequencies listed in Table~\ref{polar_parameters}. This is likely because the source appears to be more extended at this frequency and shows high polarization in the peripheral regions. When comparing the same region, the percentages are similar at 1.4 and 4.7 GHz and consistent with a constant value of $\sim$ 35\%.

\section{Discussion}

\subsection{Source general properties}
The general properties of \emi can be summarized as follows:

1) It is a giant radio source coincident with a poor galaxy group in the supercluster SCL245 (see the Introduction). The brightest galaxy member (BGG1) is a LEG located at the center of \emi and coincident with a discrete radio source. The mass of the group is low and is consistent with the lack of detection in the X-ray band. The central group galaxies may be in a coalescence phase. 

2) Its largest linear size in the radio band is 1.5 Mpc in the east-west direction. In the center, an extended substructure is present with a slightly higher surface brightness, oriented at PA 40$^\circ$ and with a size of $\sim$ 430 Kpc.  

3) The source is well resolved in the direction perpendicular to the largest structure. The transverse size is $\sim$ 0.5 Mpc and remains approximately constant across most of the source structure. In the LOFAR image (see Fig.~\ref{compact}) the transverse size is slightly larger on the western side.

4) The total source spectral index shows a clear steepening at high frequency. The spectral-index distribution is relatively flat and uniform in the central region and steepens toward the external eastern and western regions.

5) The source age increases from BGG1 toward the peripheral regions.

6) The source is highly polarized in the L and C bands, with a similar percentage of polarized flux and PA orientation. The magnetic-field orientation is roughly parallel to the outer structure.

\subsection{A new giant radio galaxy}

Based on the optical and radio properties of the \emi region, we confirm that this source cannot be Galactic, as discussed in previous studies cited in the Introduction.

We consider identification with an optical object undetected due to a high redshift unlikely, given the large size of this radio source.

In the past, it was classified as a diffuse radio relic possibly associated with A786 or a poor galaxy group.
However, the radio properties of \emi differ significantly from those of extended diffuse sources such as radio halos and relics \citep[see e.g.,][]{feretti2012,vanweeren2019}. The total radio spectrum, the spectral index trend, and the morphology of the source are clearly inconsistent with identifying the source as a halo or relic.

The position and inner central structure favor an interpretation of \emi as a giant radio emission associated with the BGG1 galaxy; therefore, we classify this source as a giant radio galaxy (GRG). The first GRGs were discovered by \citet[][3C236 and DA240]{Willis1974} and have since raised important questions about their formation and evolution. Recently, their number has increased substantially because of the availability of high-resolution, high-sensitivity images at low frequency. \cite{Dabhade2020} and \cite{Simonte2024} used large samples of GRGs to investigate their general properties. We compare the properties of \emi with their results.

With its projected size of 1.5 Mpc, \emi is among the largest low-power GRGs. 
Figure~4b of \cite{Dabhade2020} indicates that only five sources have radio power below 10$^{25}$ W/Hz at 1.4 GHz and a size greater than 1 Mpc. The large size and long radio emitting age of \emi may be favored by the sub-microGauss magnetic-field level.

The low-frequency spectral index of \emi is significantly steeper than that of other GRGs with similar power \citep[see Fig.~4 in][]{Dabhade2020}.

\emi is one of the few GRGs with an FR I morphology. The optical galaxy BGG1, with M$_r$ = -22.4, is close to the peak of the magnitude distribution of FR I hosts in the FIRST catalog of FRI radio galaxies (FRICAT) \citep{capetti2017}. The non-detection of BGG1 in the  VLASS, as previously noted, is not inconsistent with its identification as the parent galaxy of \emi. In our highest-resolution image (see the top panel of Figure~\ref{core}), the discrete source coincident with BGG1 is clearly visible and not confused with the diffuse emission. Moreover, the spectral-index distribution shows that this region has the flattest spectrum. To study the possible radio structure of BGG1 on the kiloparsec and sub-kiloparsec scales, deeper and higher-resolution observations than VLASS are necessary.

\emi shows a uniform radio brightness distribution and a large transverse size, which is unusual for an FR I source. Moreover, the lack of a jet-like feature is quite unusual; however, similar sources are present in the literature, for example 3C 386 \citep{Leahy1991}.
A visual inspection of GRGs morphologies reveals a very similar source, J0807+740 \citep[Fig.~2 in][]{lara2001}. The FR I morphology is consistent with low radio power, optical magnitude, and     Low-Excitation Radio Galaxy (LERG) galaxy classification.  
This FR I galaxy resides in an underdense environment, namely a small galaxy group belonging to a supercluster. This is consistent with the results of \cite{Sankhyayan2024}, who find that GRGs preferentially grow in sparse regions of the cosmic web.

 A comparison between the total radio low-frequency power of \emi and the 5 GHz radio power of the discrete source identified with BGG1 fits well with the general correlation between the total radio power at 408 MHz and the central emission at 5 GHz (arcsecond scale) of radio galaxies
found by \cite{giovannini2001}.  \cite{Lara2004} applied this statistical approach to a sample of GRGs selected from the NRAO VLA Sky Survey (NVSS) (37$\%$ FR I, 55$\%$ FR II, and 8$\%$ FR I/II) and found properties consistent with those of the general sample used by \cite{giovannini2001}. They argued that GRGs do not require powerful nuclear emission and jets.
For a proper comparison, we derived the total radio power of \emi at 408 MHz by scaling the radio power at 144 MHz with its spectral index and considering the different cosmology used here and in \cite{giovannini2001} and \cite{Lara2004}.

\cite{Stuardi2020} conducted a LOFAR study of the polarization of 179 GRG observed with HPBW = 20 arcsec, drawn from the sample of \cite{Dabhade2020}, and find a detection rate of 20\%. Giant radio galaxies (GRGs) preferentially reside in very rarefied regions and experience a low level of depolarization 
between 144 MHz and 1.4 GHz. Unfortunately we cannot derive the RM, but we find little to no depolarization between 1.4 and 4.7 GHz, and, given the high brightness of the polarized emission at 1.4 GHz, we expect that \emi is among the 20\% of polarized GRGs.

\section{Conclusions}

We identify \emi as a giant radio galaxy (GRG) with a size of 1.5 Mpc and an estimated age of $\sim$ 100 Myr.

The optical parent galaxy is a bright elliptical LERG galaxy (M$_r$ = -22.4). It is the brightest member of a poor group, and its ability to originate such a large radio source is unusual. The complex radio morphology of the central region surrounding the central discrete source (see Fig.~\ref{core}), suggests a possible connection with the energy available from the collapse of this small group, together with interactions with nearby small structures, as visible in Figure~\ref{compact}.

In the BGG1 region the spectral index is low, and the radio spectrum steepens toward the external regions, consistent with the aging of relativistic emitting particles. 

The source \emi is one of a relatively small population of low-power FR I GRGs. The surface brightness decreases from the center to the peripheral regions. We find no evidence of an interaction between the radio-emitting plasma and the IGM. The radio-emitting plasma cannot be confined by the IGM, but the magnetic field could be frozen inside the structure and prevent strong source expansion.
The polarized emission is consistent with the low IGM density in this region.

Giant radio galaxies such as \emi could explain the origin of relativistic particles and magnetic fields in low-density regions, including superclusters and filaments.

\section{Data availability}

  Table~1 is only available in electronic form at the CDS via anonymous ftp to cdsarc.u-strasbg.fr (130.79.128.5) or via http://cdsweb.u-strasbg.fr/cgi-bin/qcat?J/A+A/.

\begin{acknowledgements}

We thank the Anonymous Referee for useful comments and suggestions.
NB acknowledges support from the ERC Consolidator Grant ULU 101086378.  This paper is based (in part) on data obtained with LOFAR under project code LC2\_008.
LOFAR is the Low Frequency Array designed and constructed by ASTRON. It has observing, data processing, and data storage facilities in several countries, which are owned by various parties (each with their own funding sources), and which are collectively operated by the LOFAR ERIC under a joint scientific policy. The LOFAR resources have benefited from the following recent major funding sources: CNRS-INSU, Observatoire de Paris and Université d'Orléans, France; BMFTR, MKW-NRW, MPG, Germany; Science Foundation Ireland (SFI), Department of Business, Enterprise and Innovation (DBEI), Ireland; NWO, The Netherlands; The Science and Technology Facilities Council, UK; Ministry of Science and Higher Education, Poland; The Istituto Nazionale di Astrofisica (INAF), Italy. The National Radio Astronomy Observatory is a facility of the National Science Foundation operated under cooperative agreement by Associated Universities, Inc. This research has made use of Dark Energy Spectroscopic Instrument data  (DESI DR1): the list of the funding organization and collaborating institutions can be found in
https://data.desi.lbl.gov/doc/acknowledgments/.
This research has made use of the The Pan-STARRS Survey: the list of the funding organization and collaborating institutions can be found in https://outerspace.stsci.edu/spaces/PANSTARRS/overview.
This research made use of different packages for pyhton: APLpy \citep{Robitaille2012}, Astropy \citep{Astropy2013}, NumPy \citep{vanderwalt2011} and Matplotlib \citep{Hunter2007}.

\end{acknowledgements}
\bibliographystyle{aa}
\bibliography{biblio}

\end{document}